\documentclass[aps,twocolumn,nofootinbib,superscriptaddress,preprintnumbers,pra,10pt]{revtex4-2}

\usepackage{subcaption}
\usepackage{changepage}
\usepackage{braket}
\usepackage{graphicx,amssymb,amsmath,mathtools,bm,bbm}
\usepackage{slashed}
\usepackage{simplewick}
\usepackage{xcolor}
\usepackage{color}
\usepackage{hyperref}
\usepackage{float}
\usepackage{url}
\usepackage{dcolumn}
\begin{document}
\preprint{P3H-20-082, SI-HEP-2020-35}

\title{QCD sum rules for parameters of the $B$-meson distribution amplitudes}
    
\author{Muslem Rahimi}
\affiliation{Theoretische Physik 1, Naturwissenschaftlich-Technische Fakult\"{a}t, Universit\"{a}t Siegen, D-57068 Siegen, Germany}
\author{Marcel Wald}
\affiliation{Theoretische Physik 1, Naturwissenschaftlich-Technische Fakult\"{a}t, Universit\"{a}t Siegen, D-57068 Siegen, Germany}
\email{muslem.rahimi@uni-siegen.de}
\email{marcel.wald@uni-siegen.de}
\begin{abstract}
\vspace{5mm}
    We obtain new estimates for the parameters $\lambda_{E}^2$, $\lambda_H^2$ and their ratio $\mathcal{R} = \lambda_{E}^2/\lambda_H^2$, which appear in the second moments of the $B$-meson light-cone distribution amplitudes defined in the heavy-quark effective field theory. 
    The computation is based on two-point QCD sum rules for the diagonal correlation function and includes all contributions up to mass dimension seven in the operator-product expansion. For the ratio we get $\mathcal{R} = (0.1 \pm  0.1)$ with $\lambda_H^2 = (0.15 \pm 0.05) \, \text{GeV}^2$ and $\lambda_E^2 = (0.01 \pm 0.01) \, \text{GeV}^2$.
\vspace{3mm}
\end{abstract}
    
\maketitle

\section{Introduction}
    
    Light-cone distribution amplitudes (LCDAs) are of great importance in exclusive $B$-meson decays like $B  \rightarrow \pi \pi$ or  $B \rightarrow \pi K$ in the heavy quark limit and allow for the study of $CP$-violation in weak interactions. They parametrize matrix elements of nonlocal heavy-light currents separated along the light-cone at leading order in the heavy-quark effective theory (HQET) \cite{Georgi:1990um} in terms of expansions in wave functions of increasing twist \cite{Beneke:2003pa,Grozin:1996pq}. In particular, LCDAs appear in factorization theorems such as QCD factorization \cite{Beneke:2000ry, Beneke:1999br,Beneke:2003pa}, since these amplitudes encode the nonperturbative nature of the strong interactions and are crucial in $B$-meson decay form factor computations. 
    General definitions have been obtained in \cite{Beneke:2003pa,Grozin:1996pq}. Contrary to light-meson distribution amplitudes, which also appear in factorization theorems, the properties of the $B$-meson distribution amplitudes are less known. However, they have been extensively studied recently. Their evolution equations have been investigated for the leading twist two-particle LCDA in \cite{Lange:2003ff,Bell:2013tfa,Braun:2019zhp,Braun:2019wyx,Galda:2020epp} and for higher twist amplitudes in \cite{Braun:2017liq}. 
    Moreover, the decay $B \rightarrow \gamma \ell \nu$ is of particular interest, because it provides a simple example to probe the light-cone structure of the $B$-meson. Here, the photon has a large energy compared to the strong interaction scale $\Lambda$, so QCD factorization can be used to study parameters like the inverse moment $\lambda_B$ \cite{Beneke:2011nf,Beneke:2018wjp,Shen:2018abs,Braun:2012kp,Khodjamirian:2020hob,Wang:2016qii,Wang:2018wfj}.
    
    Three-particle LCDAs have also been investigated e.g. in \cite{Grozin:1996pq,Kawamura:2001jm}, where the corresponding Mellin moments have been defined and identities between two-particle and three-particle LCDAs have been found. In general, these three-particle LCDAs occur in higher dimensional vacuum to meson matrix elements including nonlocal quark operators. But in the case of local quark operators, these matrix elements can be expressed in terms of the parameters $\lambda_{E,H}^2$, which also contribute to the second Mellin moments of the three-particle $B$-meson distribution amplitudes.
    
    These are the parameters of particular interest in this work. They
    have been first investigated by Grozin and Neubert \cite{Grozin:1996pq} within the framework of QCD sum rules \cite{Novikov:1983gd,Shifman:1978by,Shifman:1978bx}.
    All contributions to the operator-product expansion (OPE) \cite{Wilson:1969zs} in local vacuum condensates up to mass dimension five have been considered there. Up to mass dimension four, the leading order contribution is of $\mathcal{O} (\alpha_s)$, while the leading order of the mass dimension five condensate contributes at $\mathcal{O} (\alpha_s^0)$. 
   
    The extraction of these parameters is connected to a rather large uncertainty, because the sum rules turn out to be unstable with respect to the variation of the Borel parameter. Notice that such a dependence is not unexpected, since it is well known \cite{Braun:1989iv,Ball:1998sk,Nishikawa:2011qk} that higher dimensional condensates tend to give large contributions to correlation functions including higher dimensional operators. 
    
    Further study by Nishikawa and Tanaka \cite{Nishikawa:2011qk} lead to deviations from the original values for $\lambda_{E,H}^2$. These authors argued in their work that a consistent treatment of all $\mathcal{O} (\alpha_s)$ contributions should resolve the stability problem, which is related to the fact that the OPE does not converge for the parameters $\lambda_{E,H}^2$ in \cite{Grozin:1996pq}. For this analysis, they included the $\mathcal{O}(\alpha_s)$ corrections of the coupling constant $F(\mu)$ as well, which, albeit leading to good convergence of the OPE, obey large higher order perturbative corrections \cite{Broadhurst:1991fc,Penin:2001ux}. Moreover, they included as an additional nonperturbative correction the dimension six diagram of $\mathcal{O}(\alpha_s)$ in order to check the convergence of the OPE beyond mass dimension five and calculated the $\mathcal{O} (\alpha_s)$ corrections for the dimension five condensate. 
    After performing a resummation of the large logarithmic contributions, which results into more stable sum rules and into a more convergent OPE compared to \cite{Grozin:1996pq}, they obtained new estimates for the parameters $\lambda_{E,H}^2$.
    If we compare the estimates from \cite{Grozin:1996pq} and \cite{Nishikawa:2011qk} in Table \ref{tab::finalresult}, we see that the values for $\lambda_{E,H}^2$ differ by approximately a factor of three, although the ratio $\lambda_E^2/\lambda_H^2$ gives nearly the same value. 
    
    It is therefore timely to investigate new alternative sum rules which also allow for the predictions of $\lambda_{E,H}^2$. Instead of analysing a correlation function with a three-particle and a two-particle current, we consider sum rules based on a diagonal correlation function of two quark-antiquark-gluon three-particle currents. We include all leading order contributions up to  mass dimension seven. The advantage of this sum rule is that it is positive definite and hence we expect that the quark-hadron duality is more accurate compared to \cite{Grozin:1996pq,Nishikawa:2011qk}. But due to the high mass dimension of the correlation function, we see that the OPE does not show better convergence than in the nondiagonal case. Moreover, the continuum and higher excited states are dominating the sum rule. This problem will be resolved by considering combinations of the parameters $\lambda_{E,H}^2$, in particular the $\mathcal{R}$-ratio $\mathcal{R} = \lambda_E^2/\lambda_H^2$. 
    
    The paper is organized as follows: In Sec. \ref{chp: DerivationSumRule} we derive the sum rules for the parameters $\lambda_{E,H}^2$ and the sum $(\lambda_H^2 + \lambda_E^2)$. Sec. \ref{chp: Contributions} is devoted to the computation of the OPE contributions which enter the sum rules. In Sec. \ref{chp: NumericalAnalysis} we present the numerical analysis of the sum rules and state our final results for the parameters $\lambda_{E,H}^2$. Additionally, we investigate the ratio given by the quotient of these parameters. Finally, we conclude in Section \ref{chp:Conclusion}.
    
    \hspace{-1.5cm}
\section{Derivation of the QCD Sum Rules in HQET} \label{chp: DerivationSumRule}
    
    \noindent 
	In this chapter we derive the sum rules for the diagonal quark-antiquark-gluon three-particle correlation function. Before we start, the definition of the HQET parameter $\lambda_{E,H}^2$ is in order \cite{Grozin:1996pq}:
	\begin{align}
        \bra{0} {g_s \bar{q} \; \vec{\alpha} \cdot \vec{E}} \; \gamma_5 h_v\ket{\bar{B}(v)} &= F(\mu) \, \lambda_E^2 \, , \label{eq:DefLamE} \\
        \bra{0} g_s \bar{q} \;\vec{\sigma} \cdot \vec{H} \; \gamma_5 h_v \ket{\bar{B}(v)} &= i F(\mu) \, \lambda_H^2. \label{eq:DefLamH}
    \end{align}
    From a physical point of view, these quantities parametrize the local vacuum to $\bar{B}$-meson matrix elements, which contain the chromoelectric and chromomagnetic fields in HQET. The chromoelectric field is given by $E^i = G^{0i}$ and $H^i = -\frac{1}{2} \epsilon^{ijk} G^{jk}$ denotes the chromomagnetic field, with $G_{\mu \nu} = G_{\mu \nu}^{a} T^{a}$. Here, the tensor $G^{\mu \nu} = \frac{i}{g_s} [D^{\mu},D^{\nu}]$ is the field strength tensor, while $g_s$ corresponds to the strong coupling constant. Furthermore, the fields $\bar{q}$ in Eq. \eqref{eq:DefLamE} and \eqref{eq:DefLamH} indicate light quark fields, whereas the field $h_v$ denotes the HQET heavy quark field. Moreover, $v$ is the velocity of the heavy $\bar{B}$-meson. The Dirac matrices $\alpha^i$ are given by $\gamma^0 \gamma^i$ and $\sigma^i = \gamma^{i} \gamma^{5}$. In addition to that the HQET decay constant $F(\mu)$ is defined via the matrix element
    \begin{align}
        \bra{0} \bar{q} \gamma_{\mu} \gamma_5 h_v \ket{\bar{B}(v)} = i F(\mu) v_{\mu}
    \end{align}
    and can be related to the $B$($\bar{B}$)-meson decay constant in QCD up to one loop order \cite{Neubert:1991sp}:
    \begin{align}
        f_B \sqrt{m_B} = F(\mu) K(\mu) &= F(\mu) \Big[ 1 + \frac{C_F \alpha_s}{4 \pi} \Big(3 \cdot \mathrm{ln} \frac{m_b}{\mu} - 2 \Big)  \nonumber \\
        & + ... \Big] + \mathcal{O}\Big(\frac{1}{m_b}\Big). \label{eq:RelationFb}
    \end{align}
    Its explicit scale dependence has to cancel with the one of the matching prefactor in order to lead to the constant $f_B$. Values for $f_B$ can be found in \cite{Aoki:2016frl} and estimate this decay constant to be:
    \begin{align}
        f_B = (192.0 \pm 4.3) \; \mathrm{MeV} \, .
        \label{eq::physicaldecayconstant}
    \end{align}
    The coupling constant $F(\mu)$ will be of particular importance for the derivation of the relevant low-energy parameters in the following QCD sum rule analysis. But since we are investigating the sum rules at leading order accuracy, corrections of the order $\mathcal{O}(\alpha_s)$ and $\mathcal{O}\Big(\frac{1}{m_b}\Big)$ will be neglected.
    
    As already discussed before, Grozin and Neubert \cite{Grozin:1996pq} introduced the parameters $\lambda_{E,H}^2$. For this, they considered the correlation function shown in Eq. (\ref{eq:CorrFuncOffDiag}). The starting point for our calculation is the correlation function given in Eq. (\ref{eq:CorrelationFunc}).
    
     \begin{align}
        \Pi_{\text{GN}} =& \; i \int \mathrm{d}^d x e^{-i \omega v \cdot x} \bra{0} T\{\bar{q}(0) \Gamma_1^{\mu \nu} g_s G_{\mu \nu}(0) h_v(0) \nonumber \\ &\times \bar{h}_v(x) \gamma_5 q(x) \} \ket{0} \, , \label{eq:CorrFuncOffDiag} \\
        \Pi_{\text{diag}} =& \; i \int \mathrm{d}^d x \; e^{-i \omega v \cdot x} \bra{0} T\{\bar{q}(0) \Gamma_1^{\mu \nu} g_s G_{\mu \nu}(0) h_v(0) \nonumber \\ &\times \bar{h}_v(x) \Gamma_2^{\rho \sigma} g_s G_{\rho \sigma}(x) q(x)\} \ket{0} \, . \label{eq:CorrelationFunc}
    \end{align}
	\noindent
	 Notice that at this point we do not require a specific choice of the quantities $\Gamma_1^{\mu \nu}$ and $\Gamma_2^{\rho \sigma}$, which indicate an arbitrary combination of Dirac $\gamma$-matrices, but in the following steps it is convenient to choose these matrices such that combinations of the HQET parameters $\lambda_{E,H}^2$ are projected out. This requires that the perturbative and nonperturbative contributions to the OPE in Sec. \ref{chp: Contributions} are computed for general $\Gamma_1^{\mu \nu}$ and $\Gamma_2^{\rho \sigma}$. Since we are considering a diagonal Greens function, the structure of $\Gamma_2^{\rho \sigma}$ is directly related to $\Gamma_1^{\mu \nu}$ by replacing indices. 
	 From now on we use the notation:
	 \begin{align}
	     \Gamma_1 &\equiv \Gamma_1^{\mu \nu} \, , \\
	     \Gamma_2 &\equiv \Gamma_2^{\rho \sigma} \, .
	 \end{align}
	
	Moreover, we are working in the $\bar{B}$-meson rest frame, where $ v = (1,\vec{0})^T$, in order to simplify the calculations.
	
	The next step in the derivation of the sum rules will be to exploit the unitary condition, where the ground state $\bar{B}$-meson is separated from the continuum and excited states: \\
	\begin{widetext}
	\begin{align}
		\frac{1}{\pi} \mathrm{Im} \Pi_{\text{diag}}(\omega) =& \sum_n (2\pi)^3 \delta(\omega - p_n) \bra{0} \bar{q}(0) \Gamma_1 g_s G_{\mu \nu}(0) h_v(0)\ket{n} \bra{n} \bar{h}_v(x) \Gamma_2 g_s G_{\rho \sigma}(x) q(x) \ket{0} \mathrm{d} \Phi_n \nonumber \\ =& \, \; \delta(\omega - \bar{\Lambda}) \bra{0} \bar{q}(0) \Gamma_1 g_s G_{\mu \nu}(0) h_v(0)\ket{\bar{B}} \bra{\bar{B}} \bar{h}_v(0) \Gamma_2 g_s G_{\rho \sigma}(0) q(0) \ket{0} \nonumber \\& \; + \rho^{\text{hadr.}}(\omega) \Theta(\omega - \omega^{th}) \, . \label{eq:UnitarityCond}
	\end{align}
	\end{widetext} 
	In Eq. (\ref{eq:UnitarityCond}), we introduced the binding energy $\bar{\Lambda} = m_B - m_b$, which is one of the important low-energy parameters in this formalism. Furthermore, we separated the full $n$-particle contribution in the first line into a ground state contribution, which will be the dominant contribution in our chosen stability window, and a continuum contribution including broad higher resonances. 
	In the case of QCD correlation functions, the exponential in Eq. \eqref{eq:CorrelationFunc} would generally take the form $e^{-iqx}$ with $q$ denoting the external momentum. Due to the fact that there is no spatial component in the $B$-meson rest frame, transitions from the ground state to the excited states in Eq. \eqref{eq:UnitarityCond} are possible by injecting energy $q^0$ into the system. In this work we explicitly chose $q = \omega \cdot v$ such that we end up with the correlation function shown in Eq. \eqref{eq:CorrelationFunc}.
	
	The matrix elements occurring in \eqref{eq:UnitarityCond} can be decomposed in the following way \cite{Grozin:1996pq,Nishikawa:2011qk}:
	\begin{align}
		& \bra{0} \bar{q}(0) \Gamma_1 g_s G_{\mu \nu}(0) h_v(0) \ket{\bar{B}} = \; \frac{-i}{6} F(\mu) \{\lambda_H^2(\mu) \nonumber \\
		& \times \mathrm{Tr}[\Gamma_1 P_+ \gamma_5 \sigma_{\mu \nu}]  + [\lambda_H^2(\mu) - \lambda_E^2(\mu)] \nonumber \\
		& \times \mathrm{Tr}[\Gamma_1 P_+ \gamma_5 (i v_{\mu} \gamma_{\nu} - i v_{\nu} \gamma_{\mu})]\}. \label{eq:Decomp}
	\end{align}
	
	Notice that the second decomposition is indeed valid since the $B$-meson ground state explicitly depends on the velocity $v$ and $\sigma_{\mu \nu} = \frac{i}{2} [\gamma_{\mu}, \gamma_{\nu}]$ corresponds to the usual antisymmetric Dirac tensor. In \eqref{eq:Decomp} we made use of the covariant trace formalism, further investigated in \cite{Grozin:1996pq,Falk:1992fm}. 
	
	The next step will be to use the standard dispersion relation, after using the residue theorem and the Schwartz reflection principle \footnote{For more details on QCD sum rules or HQET sum rules, see \cite{Neubert:1993mb, Colangelo:2000dp}}: \\
	\begin{align}
		\Pi_{\text{diag}}(\omega) =& \; \frac{1}{\pi} \int_0^{\infty} \mathrm{d}s \frac{\mathrm{Im} \Pi_{\text{diag}}(s)}{s - \omega - i0^+} \nonumber \\ &= \frac{1}{\bar{\Lambda} - \omega - i0^+} \bra{0} \bar{q}(0) \Gamma_1 g_s G_{\mu \nu}(0) h_v(0)\ket{\bar{B}} \nonumber \\
		& \times \bra{\bar{B}} \bar{h}_v(0) \Gamma_2 g_s G_{\rho \sigma}(0) q(0) \ket{0} \nonumber \\& \; + \int_{s^{th}}^{\infty} \mathrm{d} s \frac{\rho^{\text{hadr.}}(s)}{s - \omega - i0^+} \, .
		\label{eq::ground-higher}
	\end{align} \\
    In Eq. (\ref{eq::ground-higher}) we introduce the threshold parameter $s^{th}$, which is another relevant low-energy parameter that separates the ground state contribution from higher resonances and continuum contributions.
    
	We can now move on and evaluate the ground state contribution: 

\begin{widetext}
\begin{align}
		\bra{0} \bar{q}(0) \Gamma_1 g_s G_{\mu \nu}(0) h_v(0)\ket{\bar{B}} \bra{\bar{B}} \bar{h}_v(0) \Gamma_2 g_s G_{\rho \sigma}(0) q(0) \ket{0}  &= \; \frac{-i}{6} F(\mu) \Big[\lambda_H^2(\mu) \mathrm{Tr}[\Gamma_1 P_+ \gamma_5 \sigma_{\mu \nu}] \nonumber \\
		& + [\lambda_H^2(\mu) - \lambda_E^2(\mu)] \mathrm{Tr}[\Gamma_1 P_+ \gamma_5 (i v_{\mu} \gamma_{\nu} - i v_{\nu} \gamma_{\mu})]\Big] \label{eq:GroundStateContr} \\& \times \frac{-i}{6} F^{\dagger}(\mu) \Big[ \lambda_H^2(\mu)  \mathrm{Tr}[\gamma_5 P_+ \Gamma_2 \sigma_{\rho \sigma}] \nonumber  \\
		& - [\lambda_H^2(\mu) - \lambda_E^2(\mu)]  \mathrm{Tr}[\gamma_5 P_+ \Gamma_2 (i v_{\rho} \gamma_{\sigma} - i v_{\sigma} \gamma_{\rho})] \Big] \, \nonumber.
\end{align}
\end{widetext}
	Notice that the term involving the difference of both HQET parameter $(\lambda_H^2 - \lambda_E^2)$ does not change its sign under complex conjugation. 
	
	
	In order to derive the sum rules which ultimately determine the parameters $\lambda_{E,H}^2$, we make an explicit choice for the matrices $\Gamma_1$ and $\Gamma_2$ \cite{Grozin:1996pq}. Following the same approach as \cite{Grozin:1996pq}, we choose our gamma matrices $\Gamma_{1,2}$ as:
	\begin{align}
		\Gamma_{1} &= \frac{i}{2} \sigma_{\mu \nu} \gamma_5 \, 
		\label{eq:G1G2E}
	\end{align}
	to obtain the $(\lambda_{H}^2 + \lambda_{E}^2)^2$ sum rule. Furthermore, for the projection of the $\lambda_H^4$ sum rule we choose
	\begin{align}
		\Gamma_{1} &= i \Bigg( \frac{1}{2} \delta_{\alpha}^{\; \nu} - v_{\nu} v^{\alpha}\Bigg) \sigma_{\mu \alpha} \gamma_5 
		\label{eq:G1G2H}
	\end{align}
	and for $\lambda_E^4$:
	\begin{align}
	    \Gamma_{1} &= i v_{\nu} v^{\alpha} \sigma_{\mu \alpha} \gamma_5 \, .
	\end{align}
	Notice that these choices are Lorentz covariant in comparison to Eq. \eqref{eq:DefLamE} and \eqref{eq:DefLamH}. The corresponding expressions for $\Gamma_2$ can be obtained from $\Gamma_1$ by replacing $\mu \rightarrow \rho$, $\nu \rightarrow \sigma$. \\
    Using the relation in Eq. \eqref{eq:GroundStateContr}, we can obtain expressions for $\Pi_{E,H}$ and $\Pi_{HE}$: \\
	\begin{align}
		\Pi_{E,H}(\omega) =& \; F(\mu)^2 \cdot \lambda_{E,H}^4 \cdot \frac{1}{\bar{\Lambda}- \omega - i0^+} + \int_{s^{th}}^{\infty} \mathrm{d}s \frac{\rho_{E,H}^{\text{hadr.}}(s)}{s - \omega - i0^+} \label{eq:SpectralFuncH} \\
		\Pi_{HE}(\omega) =& \; F(\mu)^2 \cdot (\lambda_H^2 + \lambda_E^2)^2 \cdot \frac{1}{\bar{\Lambda} - \omega - i0^+} \nonumber \\
		& + \int_{s^{\text{th}}}^{\infty} \mathrm{d}s \frac{\rho_{HE}^{\text{hadr.}}(s)}{s - \omega - i0^+} \label{eq:SpectralFuncE}
	\end{align} \\
	Note that the threshold parameter $s^{th}$ in Eq. (\ref{eq:SpectralFuncH}) does not necessarily coincide with the threshold parameter in Eq. (\ref{eq:SpectralFuncE}).
	
	To parametrize the hadronic spectral density, we make use of the global and semilocal quark-hadron duality (QHD) \cite{Poggio:1975af, Hofmann:2003qf} in order to connect the hadronic spectral density with the spectral density which is described by the OPE \cite{Wilson:1969zs,Novikov:1983gd,Neubert:1991sp,Colangelo:2000dp}. This is the essential idea of this formalism. 
	However, power suppressed nonperturbative effects become dominant in comparison to the perturbative contribution for $- |\omega| \approx \Lambda_{\text{QCD}}$. In the QCD sum rule approach \cite{Novikov:1983gd}, these effects are parametrized in terms of a power series of local condensates as a consequence of the non-trivial QCD vacuum structure. These condensates carry the quantum numbers of the QCD vacuum. For convenience, we show explicitly in Appendix \ref{chp:Condensate} the expansion and averaging of the vacuum matrix element \eqref{eq:CorrelationFunc} in order to obtain the quark condensate $\bra{0}\bar{q}q\ket{0}$, the gluon condensate $\bra{0} G_{\mu \nu}^{a} G_{\rho \sigma}^{a} \ket{0}$, the quark-gluon condensate $\bra{0}\bar{q}g_s \sigma \cdot G q\ket{0}$ and the triple-gluon condensate $\bra{0} g_{s}^3 f^{a b c} G_{\mu \nu}^{a} G_{\rho \sigma}^{b} G_{\alpha \lambda}^{c} \ket{0}$. 
	
	Although we can handle the Euclidean region, the physical states described by the spectral function in Eq. \eqref{eq:SpectralFuncH} and \eqref{eq:SpectralFuncE} are defined for $\omega \in \mathbb{R}$. But since there is no estimate for the hadronic spectral density $\rho_{X}^{\text{hadr.}}(s)$, we need to make use of two statements. First, we exploit that for $\omega \ll 0$ the hadronic and the OPE spectral functions coincide at the global level:
	\begin{align} \Pi_{X}^{\text{hadr}.} = \Pi_{X}^{\text{OPE}} \; \; \; \; \text{for} \, \, \, X \in \{H,E,H\hspace{-0.1cm}E\}. \label{GlobalQHD} \end{align} 
	Asymptotic freedom guarantees that this equality holds. Moreover, we need to employ the semilocal quark-hadron duality, which connects the spectral densities:
	\begin{align} \int_{s_{X}^{th}}^{\infty} \mathrm{d}s \frac{\rho_{X}^{\text{hadr.}}(s)}{s - \omega - i0^+} = \int_{s_{X}^{th}}^{\infty} \mathrm{d}s \frac{\rho_{X}^{\text{OPE}}(s)}{s - \omega - i0^+}, \label{LocalQHD} 
	\end{align}  
	where $X$ needs be chosen according to \eqref{GlobalQHD}. 
	In the low-energy region, where nonperturbative effects dominate, the duality relation is largely violated due to strong resonance peaks, while in the high-energy region these peaks become broad and overlapping. Once a sum rule is obtained, the approximations made by QHD are consistent (see Section \ref{chp: NumericalAnalysis} for more details). So it is necessary to work in the transition region where the condensates are important, but still small and local enough such that perturbative methods can be applied.

    Based on the relations in Eq. \eqref{GlobalQHD}, \eqref{LocalQHD}, we separate the integral over the OPE spectral density by introducing the threshold parameter $s^{th}$. Hence, we end up with the following form for the sum rules:
	\begin{align}
		F(\mu)^2 \cdot \lambda_{E,H}^4 \frac{1}{\bar{\Lambda} - \omega - i0^+} =& \int_{0}^{s^{th}} \mathrm{d}s \frac{\rho_{E,H}^{\text{OPE}}(s)}{s - \omega - i0^+} \, \label{eq::hadronrep_1}, \\
		F(\mu)^2 \cdot (\lambda_H^2 + \lambda_E^2)^2 \frac{1}{\bar{\Lambda} - \omega - i0^+} =& \int_{0}^{s^{th}} \mathrm{d}s \frac{\rho_{HE}^{\text{OPE}}(s)}{s - \omega - i0^+} \label{eq::hadronrep_2}.
	\end{align} \\
	Finally, we perform a Borel transformation, which removes possible subtraction terms and leads further to an exponential suppression of higher resonances and the continuum. In addition to that, the convergence of our sum rule is improved. Generally, the Borel transform can be defined in the following way \cite{Neubert:1993mb,Colangelo:2000dp}:
	\begin{align}
	    \mathcal{B}_M f(\omega) = \underset{n \rightarrow \infty, - \omega \rightarrow \infty}{\mathrm{lim}} \frac{(-\omega)^{n + 1}}{\Gamma(n + 1)} \Big(\frac{\mathrm{d}}{\mathrm{d} \omega} \Big)^n f(\omega),
	\end{align}
	where $f(\omega)$ illustrates an arbitrary test function. Furthermore, we keep the ratio $M = \frac{-\omega}{n}$ fixed, $M$ denotes the Borel parameter.
	
	After applying this transformation, we derive the final form of our sum rule expressions:
	\begin{align}
		F(\mu)^2 \cdot \lambda_{E,H}^4 \cdot e^{-\bar{\Lambda}/M} &= \int_{0}^{\omega_{th}} \mathrm{d} \omega \;  \rho_{E,H}^{\text{OPE}}(\omega) \; e^{-\omega/M} \nonumber \\
		& = \int_{0}^{\omega_{\text{th}}} \mathrm{d} \omega \; \frac{1}{\pi} \mathrm{Im} \Pi_{E,H}^{\text{OPE}}(\omega) \; e^{-\omega/M} \, , 
        \label{eq:SumRuleH}
	\end{align} 
	\begin{align}
		F(\mu)^2 \cdot (\lambda_H^2 + \lambda_E^2)^2 \cdot e^{-\bar{\Lambda}/M} &= \int_{0}^{\omega_{th}} \mathrm{d} \omega \;  \rho_{HE}^{\text{OPE}}(\omega) \; e^{-\omega/M} \nonumber \\
		& \hspace{-0.5cm} =  \int_{0}^{\omega_{\text{th}}} \mathrm{d}\omega \; \frac{1}{\pi} \mathrm{Im} \Pi_{HE}^{\text{OPE}}(\omega) \; e^{-\omega/M} \, .\label{eq:SumRuleE}
	\end{align}
	These are the QCD sum rules presented in the paper. In order to obtain reliable values for the parameters $\lambda_{E,H}^2$ from the sum rules in Eq. \eqref{eq:SumRuleH} and \eqref{eq:SumRuleE}, the Borel parameter $M$ needs to be chosen accordingly together with the threshold parameter $\omega_{th}$. The next step will be to determine the spectral function $\Pi_{X}^{\text{OPE}}(s)$, which is given by the OPE: \\
	\begin{align}
	    \Pi_{\text{X}}^{\text{OPE}}(\omega) =& \; C^{\text{X}}_{\text{pert}}(\omega) + C^{\text{X}}_{\bar{q}q} \braket{\bar{q}q} + C^{\text{X}}_{G^2} \braket{\frac{\alpha_s}{\pi} G^2} \nonumber \\
	    & + C^{\text{X}}_{\bar{q}Gq} \braket{\bar{q} g_s \sigma \cdot G q}  + C^{\text{X}}_{G^3} \braket{g_s^3 f^{abc} G^{a} G^{b} G^{c}} \nonumber \\ &  + C^{\text{X}}_{\bar{q}qG^2} \braket{\bar{q}q} \braket{\frac{\alpha_s}{\pi} G^2} + ... \label{eq:OPE}
	\end{align}
    The Wilson coefficients $C$ in Eq. \eqref{eq:OPE} will be discussed in Sec. \ref{chp: Contributions}. Moreover, we define a more convenient notation for the condensate contributions:
    \begin{align}
        & \braket{\bar{q}q} := \bra{0} \bar{q} q \ket{0}, \braket{G^2} := \bra{0} G_{\mu \nu}^a G^{a, \mu \nu} \ket{0}, \nonumber \\
        & \braket{\bar{q} g_s \sigma \cdot G q} := \bra{0} \bar{q} g_s G^{\mu \nu} \sigma_{\mu \nu} q  \ket{0}, \nonumber \\ & \braket{g_s^3 f^{abc} G^{a} G^{b} G^{c}} := \bra{0} g_s^3 f^{abc} G^{a}_{\mu \nu} G^{b,\nu \rho} G^{c,\mu}_{\rho} \ket{0}.
    \end{align}
    
    \noindent
    As previously mentioned, the condensates are uniquely parametrized up to mass dimension five. Starting at dimension six and higher, there occur many different possible contributions, but some of them are related by QCD equations of motions and Fierz identities \cite{Thomas:2007gx} to each other \footnote{A list is given for example in the review \cite{Gubler:2018ctz}.}. Note that in the power expansion of Eq. \eqref{eq:OPE} we have only stated the dimension six and seven condensates, which give a leading order contribution to the parameters $\lambda_{E,H}^2$. 
   
   Moreover, there are many estimates for the values of the condensates given in the literature, which have been obtained from e.g. lattice QCD, sum rules \cite{Gubler:2018ctz}, but obtaining values for condensates of dimension six and higher is an ongoing task due to the mixing with lower dimensional condensates. Because of the lack of these values, the vacuum saturation approximation \cite{Shifman:1978bx} is exploited in many cases, where a full set of intermediate states is introduced into the higher dimensional condensate and the assumption is used that only the ground state gives a dominant contribution. Thus, the higher dimensional condensate will be effectively reduced to a combination of lower dimensional condensates \footnote{This has already been done for the dimension seven condensate in Eq. \eqref{eq:OPE}.}. 

	\section{Computation of the Wilson Coefficients} \label{chp: Contributions}
	
	In this chapter, the leading perturbative and nonperturbative contributions to the correlation function in \eqref{eq:SpectralFuncH} and \eqref{eq:SpectralFuncE} are calculated up to dimension seven. Since the leading order of the diagonal correlator of two three-particle currents is of  $\mathcal{O}(\alpha_s)$ in the strong coupling constant, we only investigate contributions up to this order in perturbation theory. 
	For the perturbative contribution we choose the Feynman gauge for the background field, while the nonperturbative contributions to the OPE are evaluated in the fixed-point or Fock-Schwinger (FS) gauge \cite{Fock:1937dy,Schwinger:1951nm}:
	\begin{align}
	   x_{\mu} \, A^{\mu}(x) = 0 \hspace{0.5cm} \text{and} \hspace{0.5cm}  A_{\mu}(x) = \int_0^1 \mathrm{d} u \; u x^{\nu} G_{\nu \mu}(ux).
	\end{align}
	In the FS gauge, we set the reference point to $x_0 = 0$. This reference point would occur in all intermediate steps of the calculation and cancel in the end of the calculation. It is well known that this gauge is particularly useful in QCD sum rule computations. 
	
\begin{figure}[h]
	\centering
        \subfloat[]{\includegraphics[width=0.20 \textwidth]{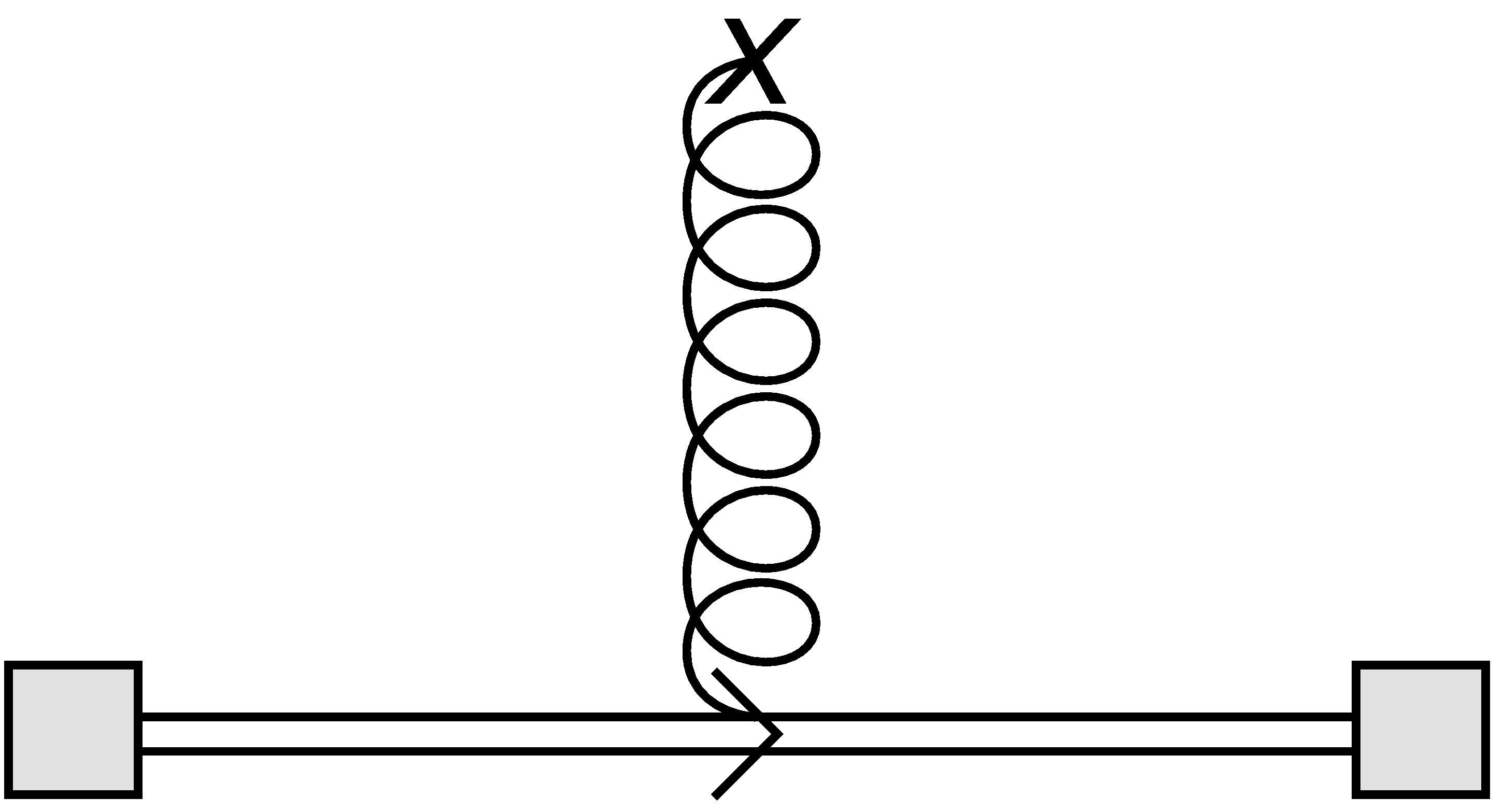}} \hspace{0.5cm}
	    \label{fig:VanishingA-contribution}
        \subfloat[]{\includegraphics[width=0.20 \textwidth]{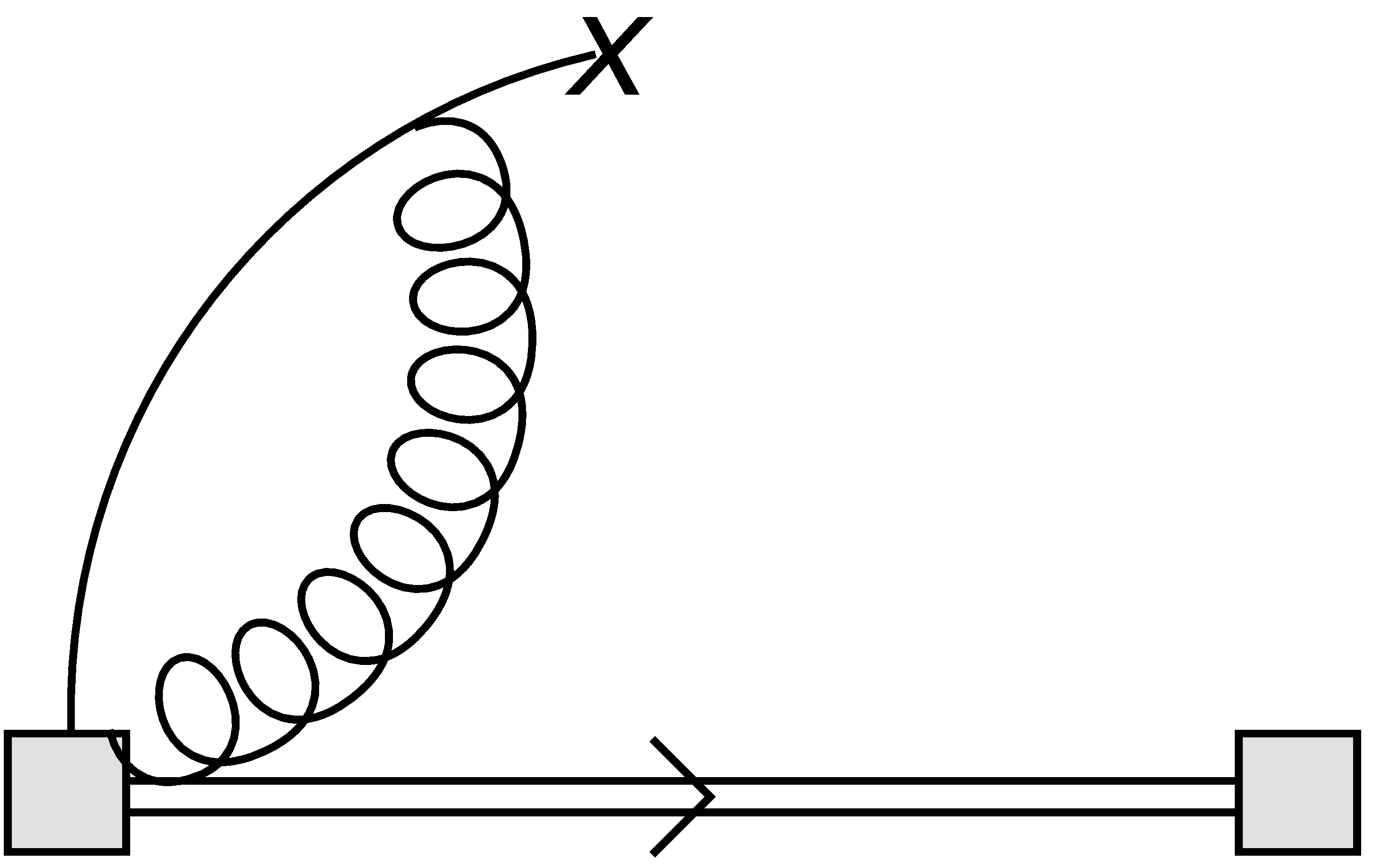}}
	    \label{fig:VanishingB-contribution}
        \hspace{0.6cm} \subfloat[]{\includegraphics[width=0.20 \textwidth]{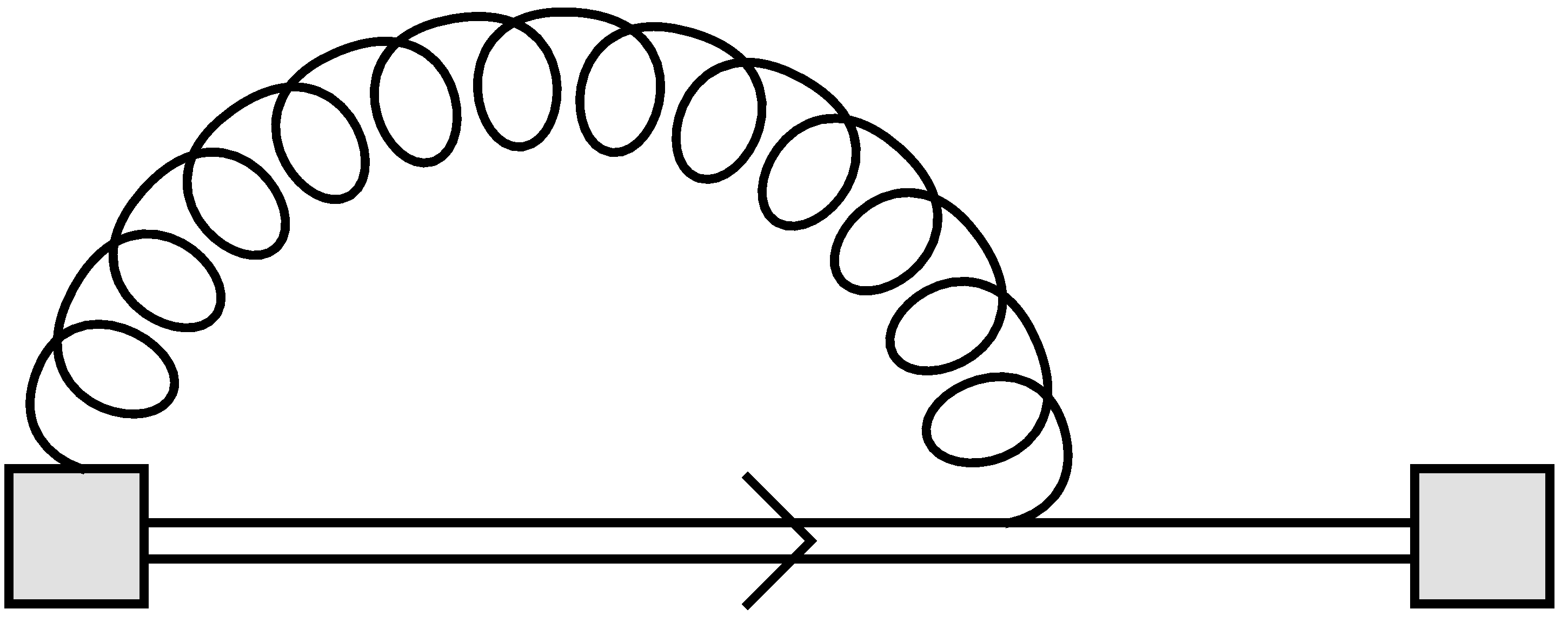} }
	    \label{fig:VanishingC-contribution}
    \caption{Vanishing subdiagrams in the Fock-Schwinger gauge.}
    \label{fig:Vanishing}
    \end{figure}

	Within the framework of QCD sum rules, the long-distance effects are encoded in local vacuum matrix elements of increasing mass dimension. In order to obtain these local condensates, the gluon field strength tensor is expanded in its spacetime coordinate $x$, which results in a simple relation between the gluon field $A_{\mu}$ and the field strength tensor $G_{\mu \nu}$. Additionally, gluon fields do not interact with the heavy quark in HQET, which can be easily seen by considering the heavy-quark propagator in position space \cite{Nishikawa:2011qk}:
	\begin{align} 
	    \contraction{}{h_v(0)}{}{\bar{h}_v(x)} h_v(0) \bar{h}_v(x) &= \Theta(-v \cdot x) \, \delta^{(d - 1)}(x_{\perp}) \, P_+ \, \mathcal{P} \; \nonumber \\
	    & \times \mathrm{exp}\Bigg( i g_s \int_{v \cdot x}^0 \mathrm{d} s v \cdot A(sv) \Bigg) \, .
	    \label{eq:HeavyQuarkWick}
	\end{align}
	Here, $x_{\perp}^{\mu} = x^{\mu} - (v \cdot x) v^{\mu}$, $P_+ = (1 + \slashed{v})/2$ denotes the projection operator and $\mathcal{P}$ illustrates the path ordering operator. Besides these simplifications, there are three additional vanishing subdiagrams depicted in Fig. \ref{fig:Vanishing} due to the FS gauge.
	
\begin{figure}[h]
	\centering
        \subfloat[]{\includegraphics[width=0.20 \textwidth]{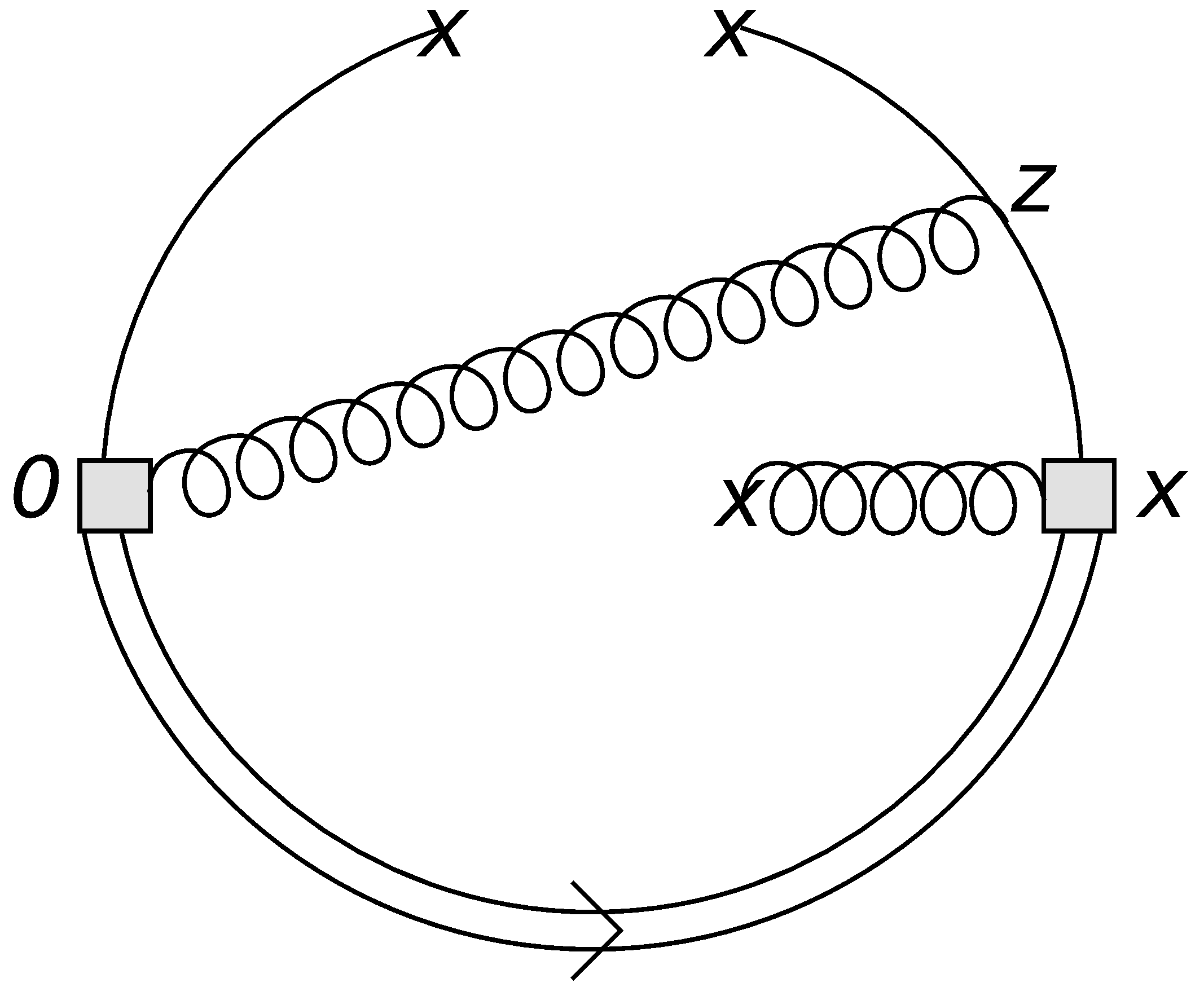}} \hspace{0.5cm}
        \label{fig:QGq1-contribution}
        \subfloat[]{\includegraphics[width=0.20 \textwidth]{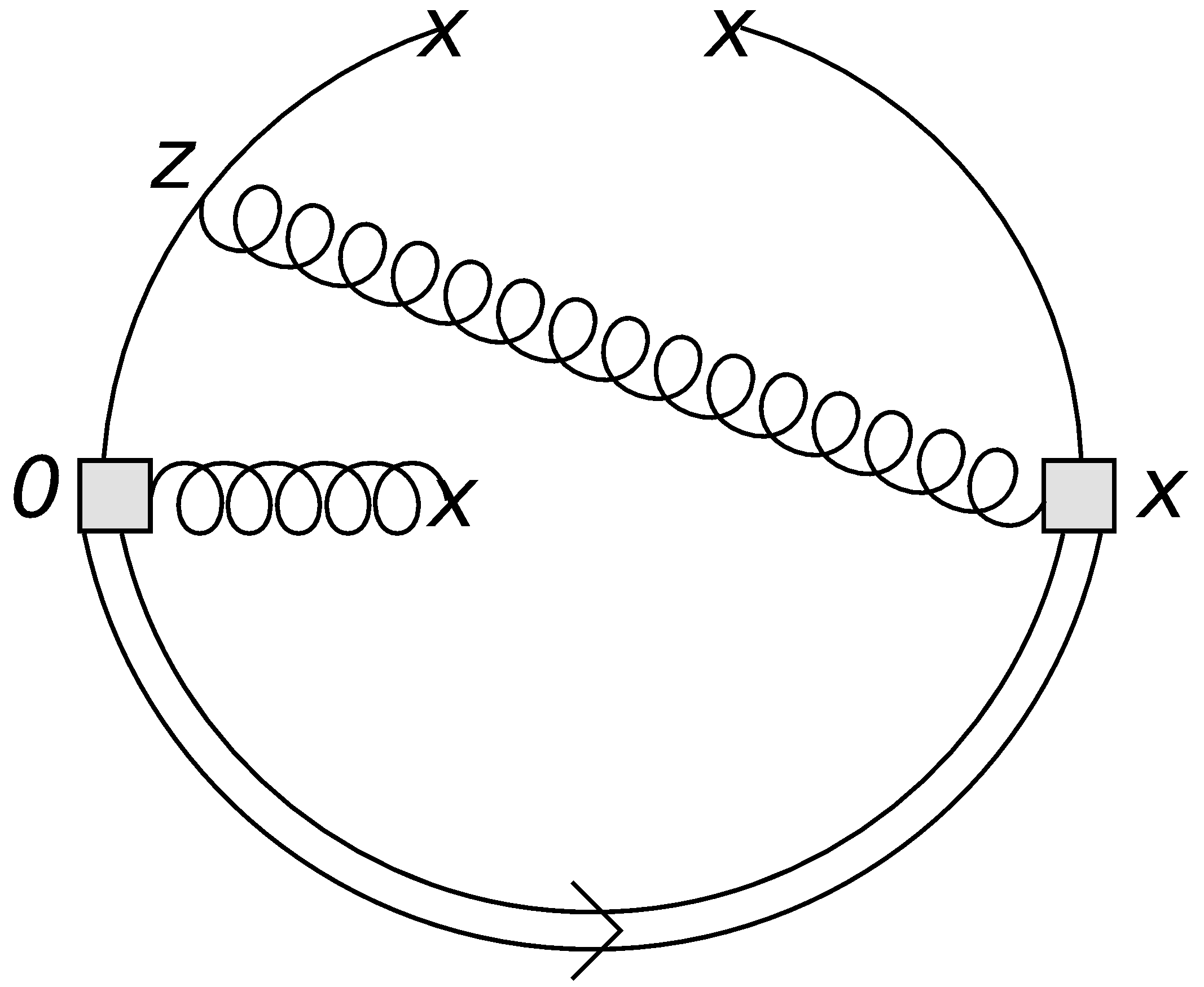}}
        \label{fig:QGq2-contribution}
        \subfloat[]{\includegraphics[width=0.20 \textwidth]{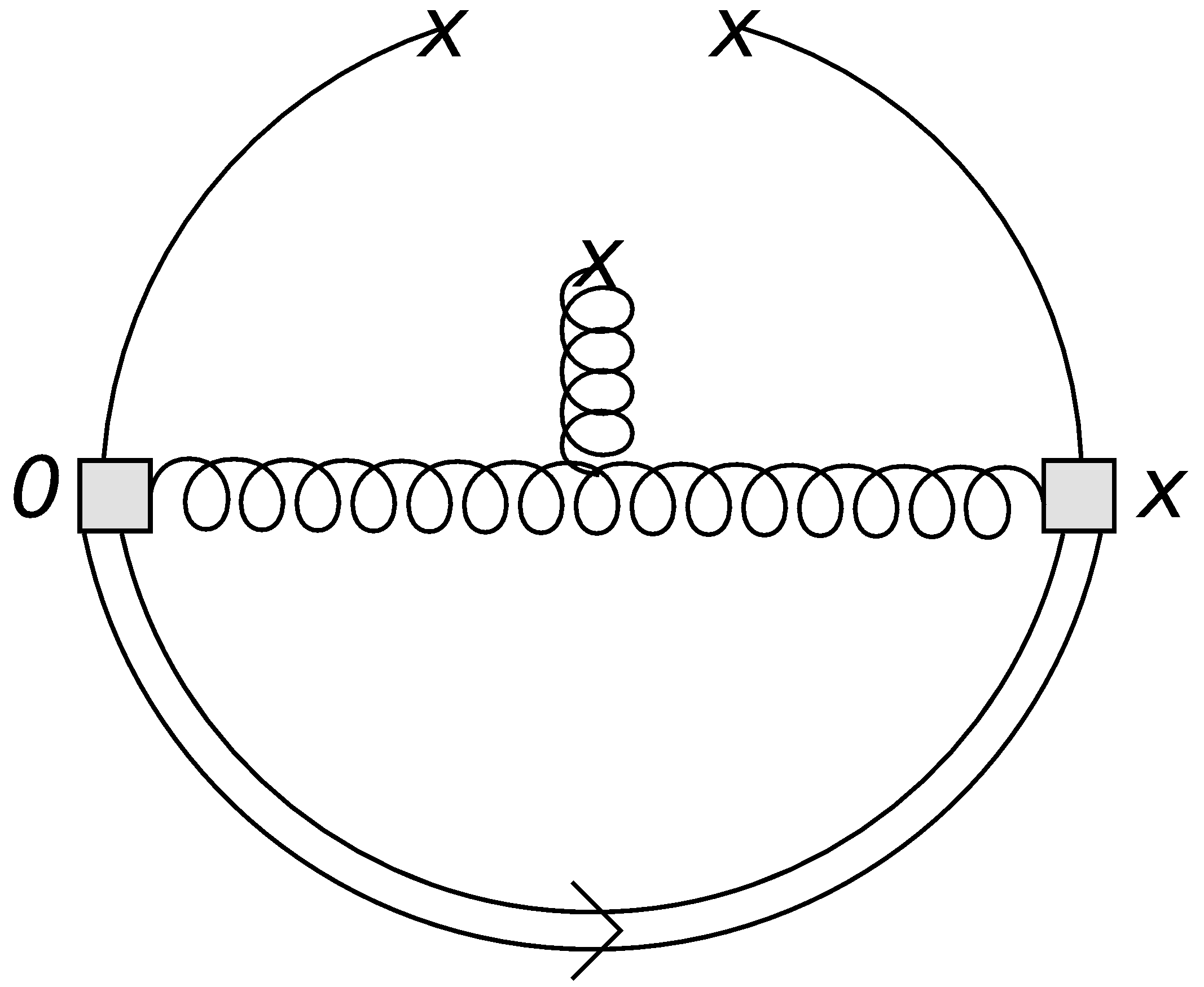}}
    \caption{Feynman diagrams for dimension five condensate contributions.}
    \label{fig:QGq-contribution}
    \end{figure}

\begin{figure}[h]
	\centering
        \subfloat[]{\includegraphics[width=0.20 \textwidth]{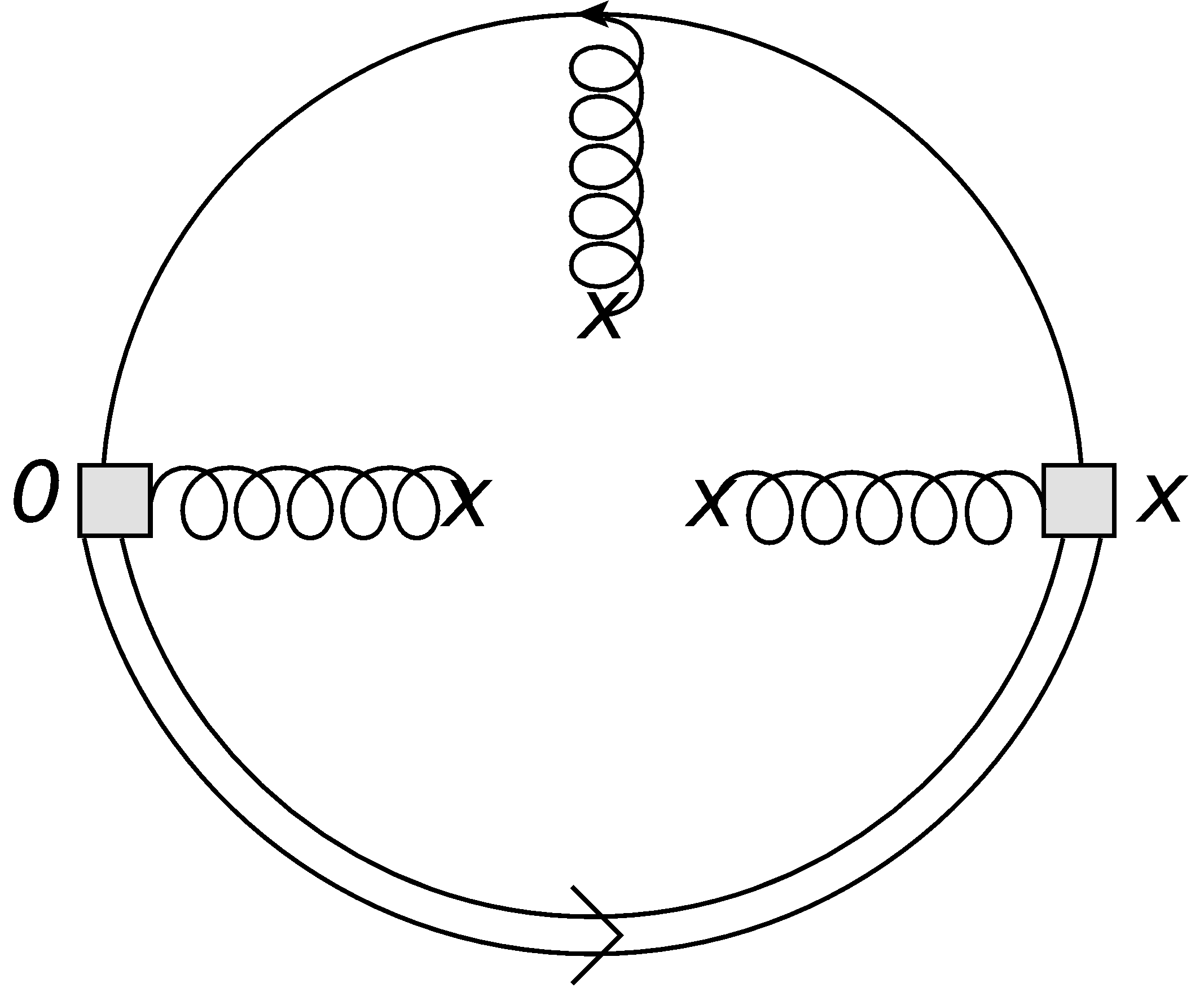}}\hspace{0.5cm}
        \subfloat[]{\includegraphics[width=0.20 \textwidth]{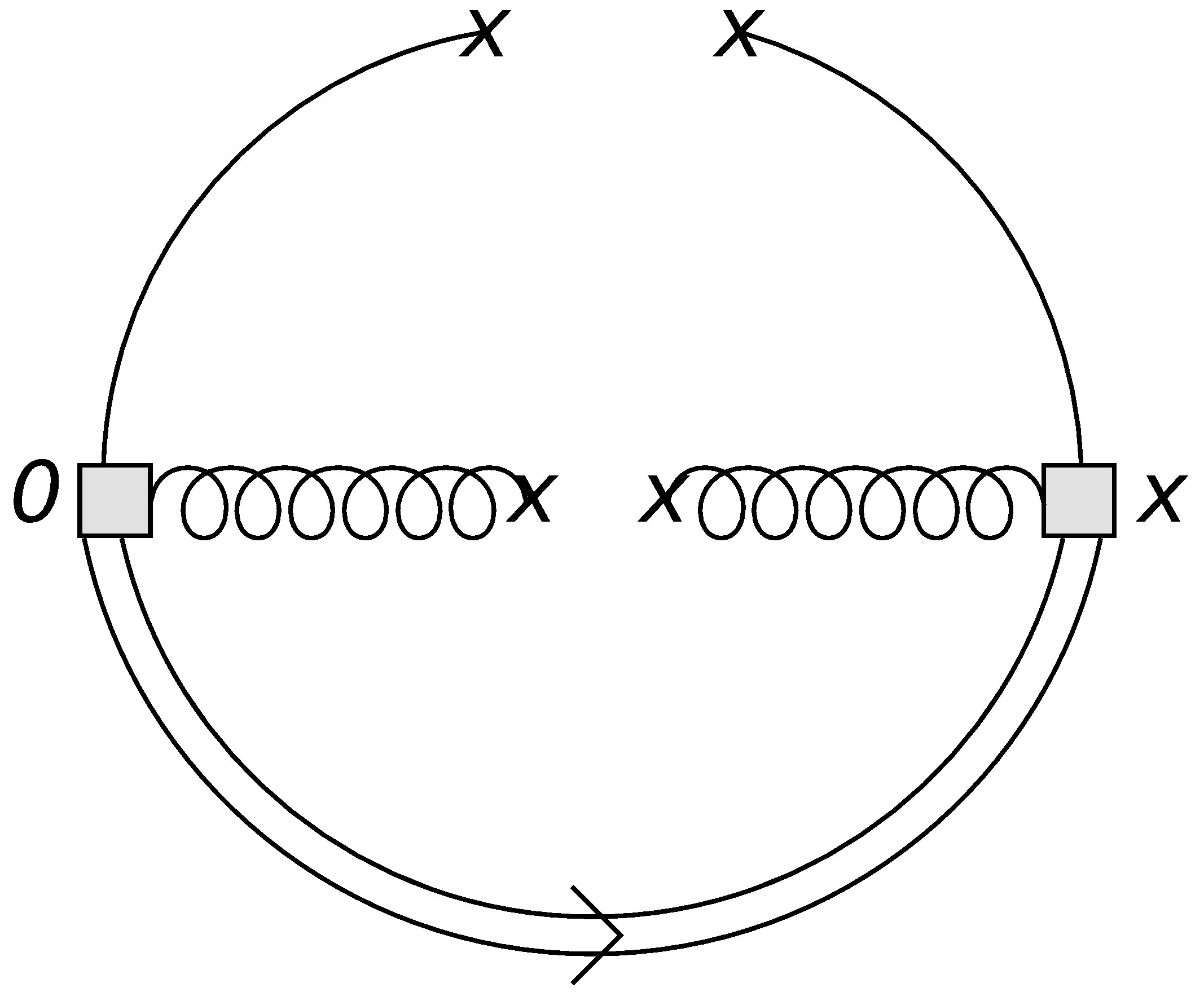}}
    \caption{Feynman diagrams for the dimension six and dimension seven condensate, which contribute to the leading order estimate of $\lambda_{E,H}^2$.}
    \label{fig:GGG&qqG2Corr-contribution}
    \end{figure}

	Generally, all diagrams can be evaluated in position space like in \cite{Grozin:1996pq,Nishikawa:2011qk}, but in this work we choose to work in momentum space. We make use of dimensional regularization for the loop integrals with the convention $d = 4 - 2 \epsilon$. 
	Fig. \ref{fig:pert&qq-contribution}-\ref{fig:Vanishing} \footnote{All diagrams in this work have been created with JaxoDraw \cite{Binosi:2008ig}.} show the diagrams which need to be computed in order to obtain the Wilson coefficients in Eq. \eqref{eq:OPE}. The calculation of these coefficients proceeds in the following way: First, we use FeynCalc \cite{Shtabovenko:2020gxv} to decompose tensor integrals to scalar integrals. In the next step, these scalar integrals are reduced to master integrals by integration-by-parts identities using LiteRed \cite{Lee:2013mka} 
	
	
	
	We start by considering the perturbative contribution and the contribution from the quark condensate in Fig. \ref{fig:pert&qq-contribution}:
	
	\begin{figure}[h]
	\centering
    \subfloat[]{\includegraphics[width = 0.20\textwidth]{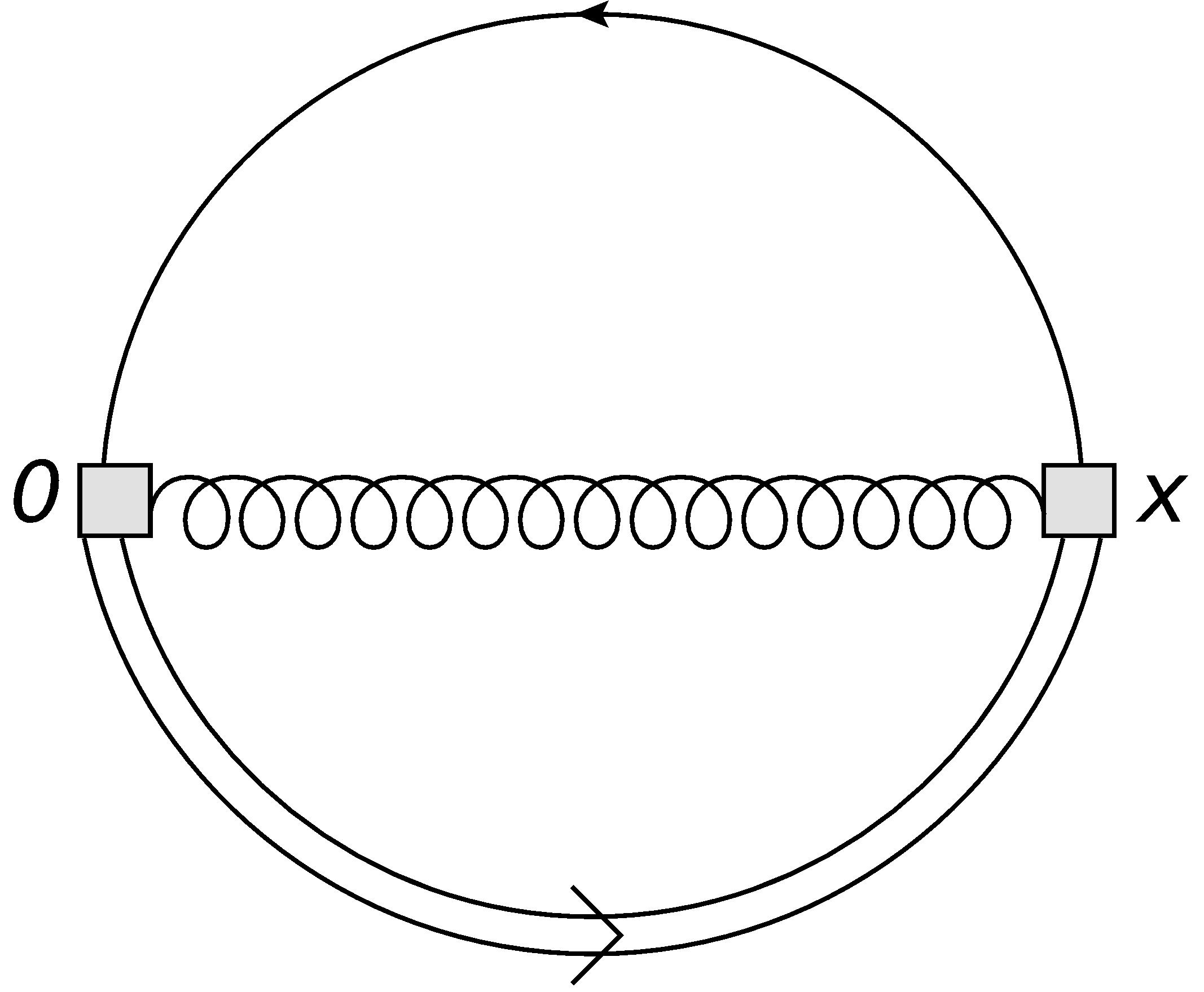}} \hspace{0.5cm}
    \subfloat[]{\includegraphics[width = 0.20\textwidth]{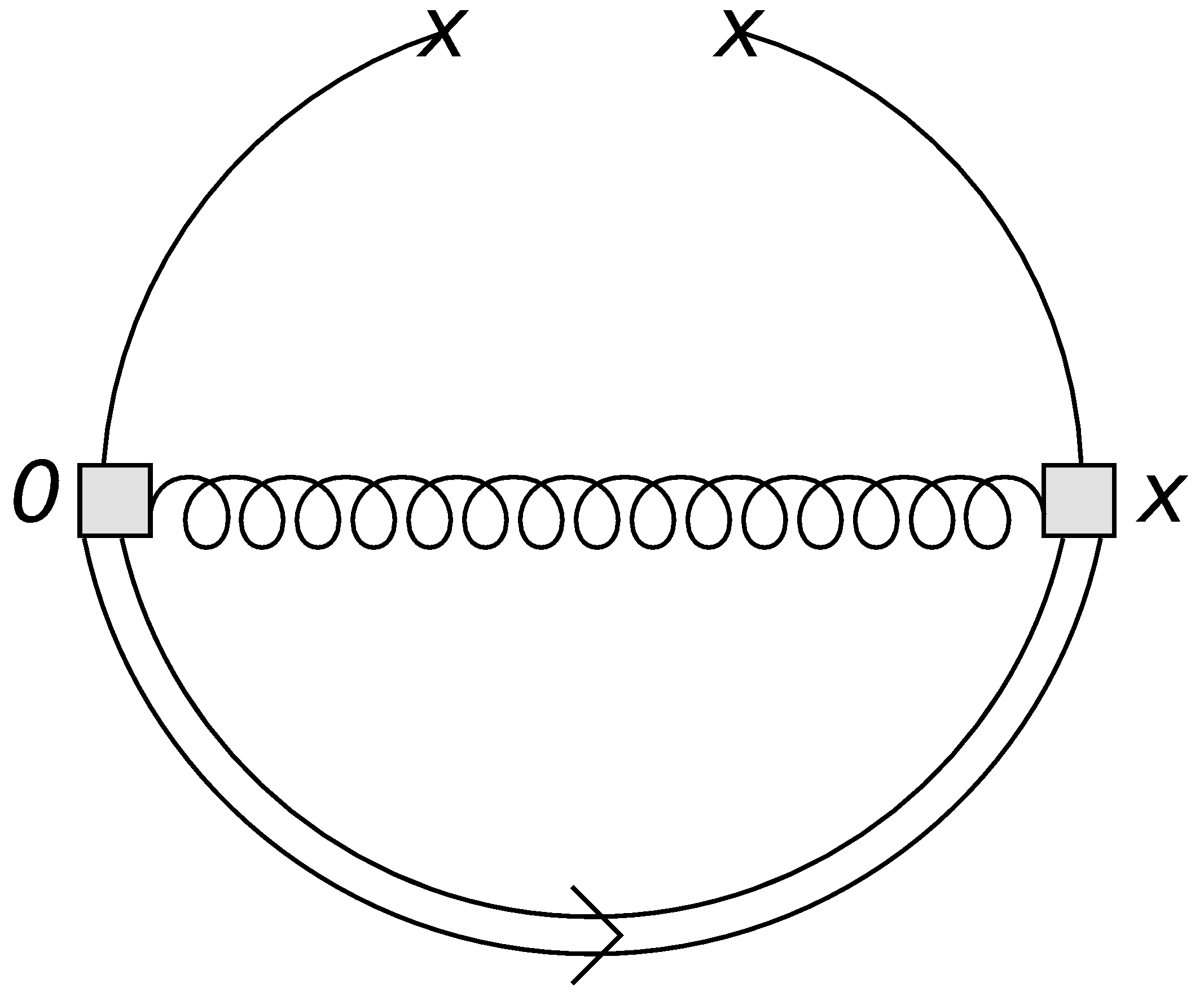}}
    \caption{Feynman diagrams for the perturbative and  $\braket{\bar{q}q}$ condensate contribution. The double line denotes the heavy quark propagator. The solid line denotes the light quark propagator and the curly line denotes the gluon propagator.}
    \label{fig:pert&qq-contribution}
    \end{figure}
    
	\begin{align}
		C^{\text{X}}_{\text{pert}}(\omega) =& \; \frac{2 \alpha_s}{\pi^3} \cdot C_F N_c \cdot \mathrm{Tr}[\Gamma_1 P_+ \Gamma_2 \slashed{v}] \cdot \bar{\mu}^{4 \epsilon}  \nonumber \\
		& \times \Gamma(-6 + 4 \epsilon) \cdot \Gamma(2 - \epsilon) \cdot \omega^{6 - 4 \epsilon} e^{4 i \pi \epsilon}  \nonumber \\& \times \Big[\Gamma(2 - \epsilon) \cdot T^1_{\mu \rho \nu \sigma} + \Gamma(3 - \epsilon) \cdot T^2_{\mu \rho \nu \sigma}\Big] \, ,
		\label{eq:WilsonCoeffPert}
	\end{align}
	\begin{align}
		C^{\text{X}}_{\bar{q}q}(\omega) &= -\frac{\alpha_s}{\pi} \cdot C_F \cdot \mathrm{Tr}[\Gamma_1 P_+ \Gamma_2] \cdot \bar{\mu}^{2 \epsilon} \cdot \Gamma(-3 + 2 \epsilon) \nonumber \\& \times \omega^{3 - 2 \epsilon} e^{2 i \pi \epsilon}  \Big[\Gamma(2 - \epsilon) \cdot T^1_{\mu \rho \nu \sigma} + \Gamma(3 - \epsilon) \cdot T^2_{\mu \rho \nu \sigma}\Big] \, ,
		\label{eq:WilsonCoeffQq}
	\end{align}
	with
	\begin{align}
	    \bar{\mu}^2 :=& \; \frac{\mu^2 e^{\gamma_E}}{4} \, , \\
	    T^1_{\mu \rho \nu \sigma} :=& \; g_{\mu \rho} g_{\nu \sigma} - g_{\mu \sigma} g_{\nu \rho} \, , \\
	    T^2_{\mu \rho \nu \sigma} :=& \; -g_{\nu \sigma} v_{\mu} v_{\rho} + g_{\mu \sigma} v_{\nu} v_{\rho} + g_{\nu \rho} v_{\mu} v_{\sigma} - g_{\mu \rho} v_{\nu} v_{\sigma} \, .
	\end{align}
	Notice that the tensor structures of $T^{1,2}_{\mu \rho \nu \sigma}$ satisfy the symmetries imposed by the field strength tensors $G_{\mu \nu}$ and $G_{\rho \sigma}$. In particular, the expressions are anti-symmetric under the exchange of $\{\mu \leftrightarrow \nu\}$, $\{\rho \leftrightarrow \sigma\}$ and symmetric under the combined exchanges $\{\mu \leftrightarrow \rho, \nu \leftrightarrow \sigma \}$ and $\{\mu \leftrightarrow \nu, \rho \leftrightarrow \sigma \}$. The Wilson coefficient for the gluon condensate and higher mass dimension correction for the quark condensate share the same tensor structure as the coefficients stated in Eq. \eqref{eq:WilsonCoeffPert} and \eqref{eq:WilsonCoeffQq}. Furthermore, the mass dimension five contribution with the non-Abelian vertex in Eq. \eqref{eq:WilsonCoeffNonAbelian} and the dimension seven contribution in Eq. \eqref{eq:WilsonDim7} make use of these tensor structures as well.
  	\begin{figure}[h]
	\centering
        \subfloat[]{\includegraphics[width=0.20 \textwidth]{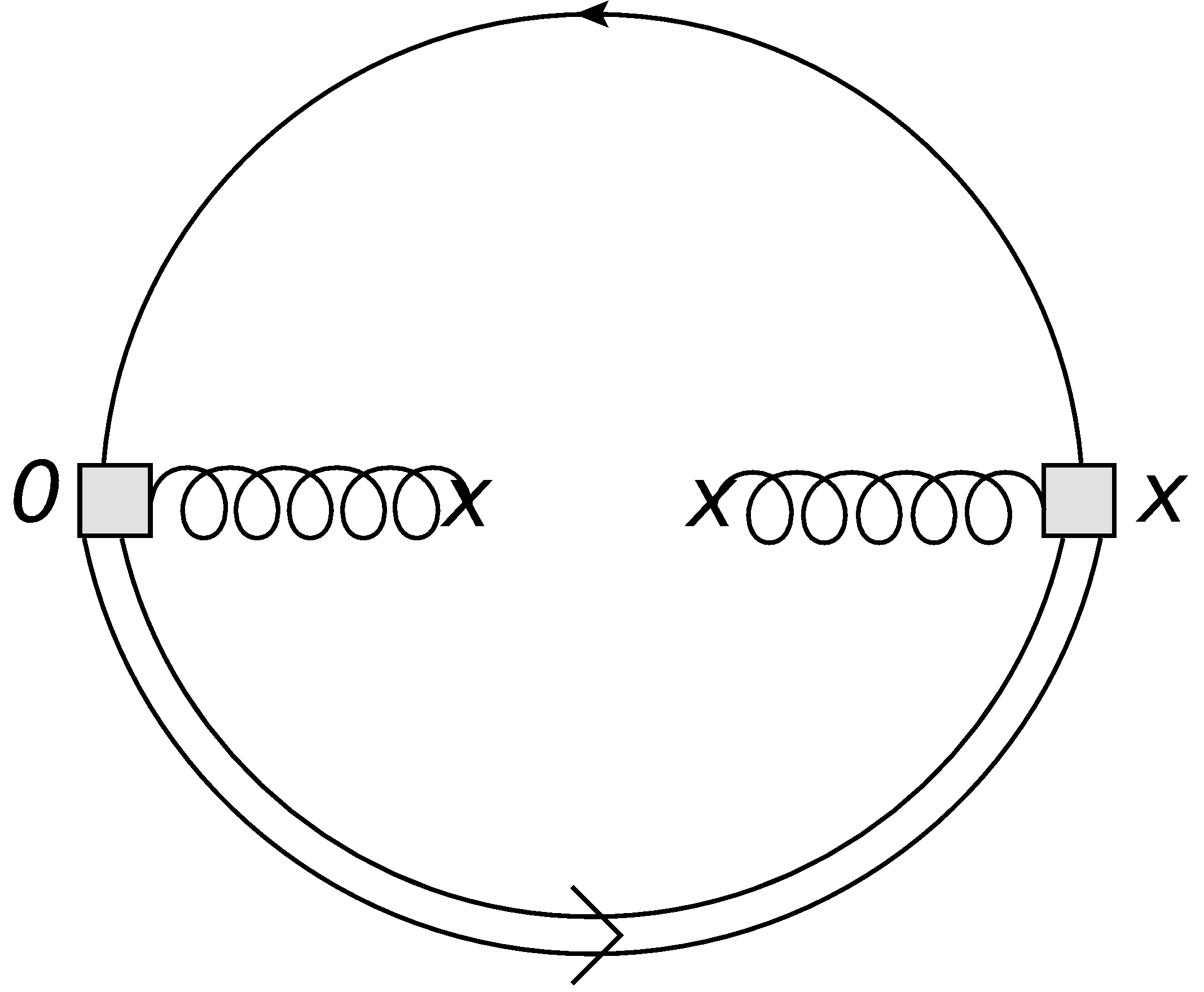}}
        \subfloat[]{\includegraphics[width = 0.20\textwidth]{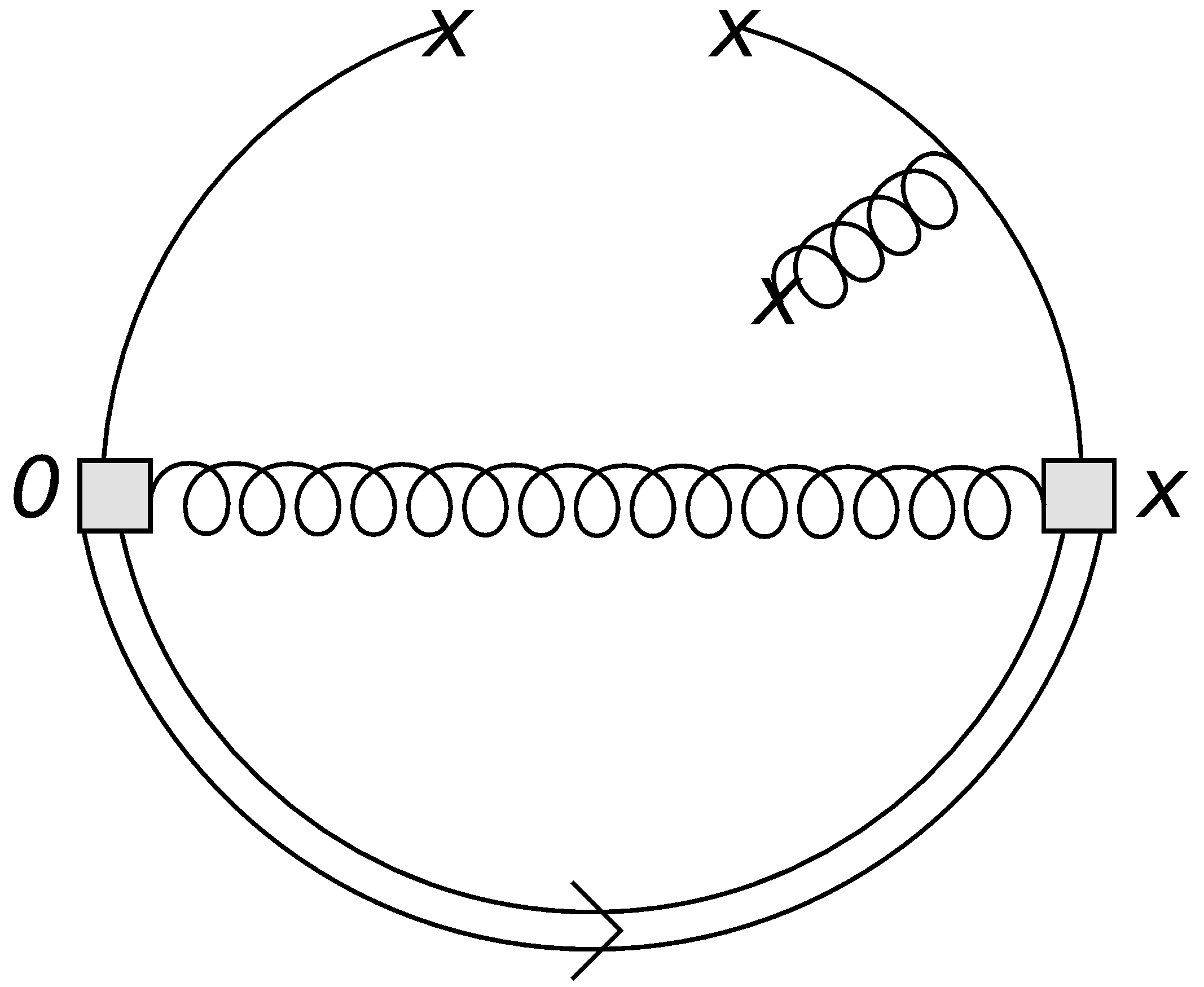}}
    \caption{(a) shows the Feynman diagram for the dimension four contribution, (b) a schematic illustration of the dimension five condensate originating from the higher order expansion of the dimension three contribution in Fig. \ref{fig:pert&qq-contribution}.}
    \label{fig:GG&qqCorr-contribution}
    \end{figure}
	
	The Wilson coefficient of the gluon condensate, which corresponds to Fig. \ref{fig:GG&qqCorr-contribution} (a) can be expressed as:
	\begin{align}
		C^{\text{X}}_{G^2}(\omega) &= \; \mathrm{Tr}[\Gamma_1 P_+ \Gamma_2 \slashed{v}] \cdot \frac{\bar{\mu}^{2 \epsilon}}{(4 - 2 \epsilon)(3 - 2 \epsilon)} \nonumber \\
		&\times \Gamma(-2 + 2 \epsilon) \cdot \Gamma(2 - \epsilon) \cdot \omega^{2 - 2 \epsilon} e^{2 i \pi \epsilon} \cdot T^1_{\mu \rho \nu \sigma} \, .
    \end{align}
	
	 The mass dimension five contributions are given as:
    \begin{align}
		C^{\text{X}}_{\bar{q}Gq,1}(\omega) =& -\frac{\alpha_s}{\pi} \cdot C_F \cdot \mathrm{Tr}[\Gamma_1 P_+ \Gamma_2] \cdot \frac{\bar{\mu}^{2 \epsilon}}{(4 - 2 \epsilon)} \nonumber \\
		& \times \Gamma(-3 + 2 \epsilon) \cdot \Gamma(3 - \epsilon) \cdot \omega^{1 - 2 \epsilon} e^{2 i \pi \epsilon} \cdot T^1_{\mu \rho \nu \sigma} \, ,
		\label{eq:Dim5QuarkExp}
	\end{align}
	
	\begin{align}
		C^{\text{X}}_{\bar{q}Gq,2}(\omega) =& \; \frac{\alpha_s}{4\pi} \cdot \frac{C_F \cdot \bar{\mu}^{2 \epsilon}}{(4 - 2 \epsilon)(3 - 2 \epsilon)} \cdot  \Gamma(-1 + 2 \epsilon) \cdot \Gamma(1 - \epsilon)  \nonumber \\
		& \times  \omega^{1 - 2 \epsilon} e^{2 i \pi \epsilon} \cdot \Big[\mathrm{Tr}[\Gamma_1 P_+ \Gamma_2 \sigma_{\mu \nu} \sigma_{\rho \sigma}]  \nonumber \\
		& - (1 - \epsilon) \cdot \mathrm{Tr}[\Gamma_1 P_+ \Gamma_2 \slashed{v} \mathrm{i} (v_{\mu} \gamma_{\nu} - v_{\nu} \gamma_{\mu}) \sigma_{\rho \sigma}] \Big] \, ,
		\label{eq:WilsonCoeffAbelian1}
    \end{align}
    \begin{align}
		C^{\text{X}}_{\bar{q}Gq,3}(\omega) &= \; \frac{\alpha_s}{4\pi} \cdot \frac{C_F \cdot \bar{\mu}^{2 \epsilon}}{(4 - 2 \epsilon)(3 - 2 \epsilon)} \Gamma(-1 + 2 \epsilon) \cdot \Gamma(1 - \epsilon) \nonumber \\
		& \times  \omega^{1 - 2 \epsilon} e^{2 i \pi \epsilon} \cdot \Big[\mathrm{Tr}[\Gamma_1 P_+ \Gamma_2 \sigma_{\mu \nu} \sigma_{\rho \sigma}] \nonumber \\
		& + (1 - \epsilon) \cdot \mathrm{Tr}[\Gamma_1 P_+ \Gamma_2 \sigma_{\mu \nu} \mathrm{i} (v_{\rho} \gamma_{\sigma} - v_{\sigma} \gamma_{\rho}) \slashed{v}] \Big] \, ,
		\label{eq:WilsonCoeffAbelian2}
	\end{align}
	\begin{align}
		C^{\text{X}}_{\bar{q}Gq,4}(\omega) &= \; \frac{i \alpha_s}{32 \pi} \cdot \frac{C_A C_F \cdot \bar{\mu}^{2 \epsilon}}{(2 - \epsilon) (3 - 2 \epsilon)} \cdot \mathrm{Tr}[\Gamma_1 P_+ \Gamma_2 \sigma^{\chi \beta}] \nonumber \\
		& \times \Gamma(-1 + 2 \epsilon) \cdot \Gamma(1 - \epsilon)  \cdot \omega^{1 - 2 \epsilon} e^{2 i \pi \epsilon} \cdot \nonumber \\& \; \Big[ \{g_{\mu \chi} T^1_{\nu \rho \beta \sigma} - (\beta \leftrightarrow \chi) \} + (1 - \epsilon) \nonumber \\
		& \times \big(\{v_{\beta} g_{\mu \rho} (v_{\sigma} g_{\nu \chi} - v_{\nu} g_{\sigma \chi}) - (\rho \leftrightarrow \sigma)\} \; + \nonumber \\& \; \{v_{\nu} g_{\mu \chi} (v_{\sigma} g_{\beta \rho} - v_{\rho} g_{\beta \sigma}) - (\beta \leftrightarrow \chi) \} \big) \Big] - (\mu \leftrightarrow \nu) \; ,  
		\label{eq:WilsonCoeffNonAbelian}
	\end{align}
	Although the other contributions for the mass dimension five condensate (Fig. \ref{fig:QGq-contribution}) possess a more complicated tensor structure, all symmetries described before are still satisfied.
	We obtain the total Wilson coefficient for the mass dimension five condensate if we sum up all four previous contributions, namely $C^{\text{X}}_{\bar{q}Gq} = \sum_{k = 1}^4 C^{\text{X}}_{\bar{q}Gq,k}$.
	The last two diagrams depicted in Fig. \ref{fig:GGG&qqG2Corr-contribution} are of mass dimension six and seven. Their contributions are expected to be smaller compared to the dimension five contributions, such that we observe that the OPE starts to converge. Other contributions to mass dimension six are vanishing or are of $\mathcal{O}(\alpha_s^2)$. Thus, the triple-gluon condensate is the only relevant condensate at this order and the Wilson coefficient reads: 
	\begin{align}
		C^{\text{X}}_{G^3}(\omega) &= \; \frac{\bar{\mu}^{2 \epsilon}}{64 \pi^2} \cdot B_{\mu \lambda \rho \nu \sigma \alpha} \cdot \Gamma(2 \epsilon) \cdot \Gamma(1 - \epsilon) \cdot \omega^{- 2 \epsilon} e^{2 i \pi \epsilon} \nonumber \\
		& \hspace{-0.5cm} \times \Big[ \mathrm{Tr}[-i \cdot \Gamma_1 P_+ \Gamma_2 \slashed{v} \sigma^{\lambda \alpha}] + \mathrm{Tr}[\Gamma_1 P_+ \Gamma_2 (v^{\alpha} \gamma^{\lambda} - v^{\lambda} \gamma^{\alpha})]]\Big] \, ,
		\label{eq:tripleGluon}
	\end{align}
	\noindent
	where the expression $B_{\mu \lambda \rho \nu \sigma \alpha}$ is defined in Appendix \ref{chp:Condensate}. Finally, we state the expression for the dimension seven contribution:
	
	\begin{align}
		C^{\text{X}}_{\bar{q} q G^2}(\omega) &= - \mathrm{Tr}[\Gamma_1 P_+ \Gamma_2] \cdot \frac{T^1_{\mu \rho \nu \sigma}}{\omega + i0^+} \cdot \frac{\pi^2}{2 N_c (4 - 2 \epsilon) (3 - 2 \epsilon)} \, .
		\label{eq:WilsonDim7}
	\end{align}
	
	According to Eq. \eqref{eq:SpectralFuncH} and \eqref{eq:SpectralFuncE}, we still need to take the imaginary part of these diagrams. We choose to compute directly the loop diagrams and take the imaginary part of the resulting expression. Following Cutkosky rules, another approach would be to perform the calculation by considering all possible cuts for the diagrams. Apart from the diagrams in Fig. \ref{fig:QGq-contribution}, the diagrams are finite. (a) and (b) in Fig. \ref{fig:QGq-contribution} include both a three-particle and a two-particle cut, where the latter requires a non-trivial renormalization procedure \cite{Grozin:1996hk}. The optical theorem states that both calculations yield the same result.
	Besides the diagram in Fig. \ref{fig:GG&qqCorr-contribution} (b), all diagrams in Fig. \ref{fig:pert&qq-contribution}-\ref{fig:GGG&qqG2Corr-contribution} can generally be calculated by using perturbative methods. Fig. \ref{fig:GG&qqCorr-contribution} (b) stems from higher order corrections in the expansion of the nonperturbative quark condensate in Eq. \eqref{eq::matrix25}.
	Moreover, the diagrams contributing to the quark-gluon condensate in Fig. \ref{fig:QGq-contribution} (a) and (b) obey the same structure as the contributions in \cite{Grozin:1996pq,Nishikawa:2011qk} and hence a cross-check is possible after replacing the quark condensate by the quark-gluon condensate and keeping in mind that the Lorentz structures differ. 
	By taking the imaginary part of all Wilson coefficients discussed above, plugging the results into Eq. \eqref{eq:SumRuleH}, \eqref{eq:SumRuleE} and performing the integration over $\omega$ up to the threshold parameter $\omega_{th}$, we obtain the final expression for the sum rules shown in Eq. \eqref{eq::LambdaHPlusE-sumrule-complete} to Eq. \eqref{eq::LambdaE-sumrule-complete}. 
	
	For convenience, we introduced the function:
	\begin{align}
	    G_n(x) := 1 - \sum_{k = 0}^n \frac{x^k}{k!} e^{-x}.
	\end{align}
	\noindent
	We see that the sum rules for $\lambda_{E,H}^4$ in Eq. \eqref{eq::LambdaH-sumrule-complete} and \eqref{eq::LambdaE-sumrule-complete} have got the same expression for the perturbative contribution. This contribution is in addition to that positive, since we are studying a positive-definite correlation function in Eq. \eqref{eq:CorrelationFunc}. Furthermore, the quark, the gluon and the triple-gluon condensate in Eq. \eqref{eq::LambdaH-sumrule-complete}, \eqref{eq::LambdaE-sumrule-complete} have different signs and the Wilson coefficients in Eq. \eqref{eq:WilsonCoeffAbelian1}, \eqref{eq:WilsonCoeffAbelian2} and \eqref{eq:WilsonCoeffNonAbelian} vanish for $\lambda_E^4$. This will have implications on the stability of the sum rule for the parameter $\lambda_E^4$ and will be investigated in Sec. \ref{chp: NumericalAnalysis}. The dimension three, four and six condensates do not appear in Eq. \eqref{eq::LambdaHPlusE-sumrule-complete}, since the signs differ in Eq. \eqref{eq::LambdaE-sumrule-complete} compared to \eqref{eq::LambdaH-sumrule-complete}.
	
	All sum rules involve the decay constant $F(\mu)$, whose calculation in terms of the correlation function can be found, e.g. in Ref. \cite{Nishikawa:2011qk}. 
	For consistency, we will retain the result at leading order in $\alpha_s$
	\begin{align}
	    F^2(\mu) \cdot  e^{-\bar{\Lambda}/M} =& \; \frac{2 N_c M^3}{\pi^2} \cdot G_2\Big(\frac{\omega_{th}}{M}\Big) \; - \braket{\bar{q} q} \nonumber \\& \; + \frac{1}{16M^2} \braket{\bar{q} g_s G \cdot \sigma q}  .
	    \label{eq:SumRuleF} 
	\end{align}
	
	\begin{widetext}
		\begin{align}
		& F(\mu)^2 \cdot (\lambda_H^2 + \lambda_E^2)^2 \, e^{-\bar{\Lambda}/M} = \; \frac{\alpha_s C_A C_F}{\pi^3} \cdot 24 M^7 \cdot G_6\Big(\frac{\omega_{th}}{M}\Big) - \frac{\alpha_s C_F C_A}{4\pi} \cdot \braket{\bar{q}g_s \sigma \cdot G q}\cdot M^2 \cdot \nonumber \\
		& \hspace{4.0cm} G_1\Big(\frac{\omega_{th}}{M}\Big) - \frac{3 \alpha_s C_F}{2\pi} \cdot \braket{\bar{q}g_s \sigma \cdot G q}\cdot M^2 \cdot G_1\Big(\frac{\omega_{th}}{M}\Big)  - \frac{\pi^2}{2 N_c} \braket{\bar{q}q} \braket{\frac{\alpha_s}{\pi} G^2} \, , \label{eq::LambdaHPlusE-sumrule-complete}\\ 
		& F(\mu)^2 \cdot \lambda_H^4 \,  e^{-\bar{\Lambda}/M} \;  = \; \frac{\alpha_s C_A C_F}{\pi^3} \cdot 12 M^7 \cdot G_6\Big(\frac{\omega_{th}}{M}\Big) - \frac{ \alpha_s C_F}{\pi} \braket{\bar{q}q} \cdot 6 \cdot M^4 \cdot G_3\Big(\frac{\omega_{th}}{M}\Big) \;  \nonumber \\
		& \hspace{3.5cm}  + \frac{1}{2} \braket{\frac{\alpha_s}{\pi} G^2} \cdot M^3 \cdot G_2\Big(\frac{\omega_{th}}{M}\Big) - \frac{\alpha_s C_F C_A}{8\pi} \cdot \braket{\bar{q}g_s \sigma \cdot G q}\cdot M^2 \cdot G_1\Big(\frac{\omega_{th}}{M}\Big) \nonumber  \\
		&\hspace{3.5cm}  - \frac{3\alpha_s C_F}{4\pi}  \cdot \braket{\bar{q}g_s \sigma \cdot G q}\cdot M^2 \cdot G_1\Big(\frac{\omega_{th}}{M}\Big) + \frac{\braket{g_s^3 f^{abc} G^a G^b G^c}}{64 \pi^2} \cdot M  \cdot \nonumber  \\
		& \hspace{3.5cm}  \; G_0\Big(\frac{\omega_{th}}{M}\Big) - \frac{\pi^2}{4 N_c} \braket{\bar{q}q} \braket{\frac{\alpha_s}{\pi} G^2} \label{eq::LambdaH-sumrule-complete} \, ,   \\
		& F(\mu)^2 \cdot \lambda_E^4 \,  e^{-\bar{\Lambda}/M} \; = \; \frac{\alpha_s C_A C_F}{\pi^3} \cdot 12 M^7 \cdot G_6\Big(\frac{\omega_{th}}{M}\Big) + \frac{ \alpha_s C_F}{\pi} \braket{\bar{q}q} \cdot 6 \cdot M^4 \cdot G_3\Big(\frac{\omega_{th}}{M}\Big) \;  \nonumber \\
		& \hspace{3.5cm} - \frac{1}{2} \braket{\frac{\alpha_s}{\pi} G^2} \cdot M^3 \cdot G_2\Big(\frac{\omega_{th}}{M}\Big) - \frac{\alpha_s C_F}{2 \pi} \cdot \braket{\bar{q}g_s \sigma \cdot G q} \cdot M^2 \cdot G_1\Big(\frac{\omega_{th}}{M}\Big) \;  \nonumber \\
		& \hspace{3.5cm} -\frac{\braket{g_s^3 f^{abc} G^a G^b G^c}}{64 \pi^2} \cdot M \cdot G_0\Big(\frac{\omega_{th}}{M}\Big) - \frac{\pi^2}{4 N_c} \braket{\bar{q}q} \braket{\frac{\alpha_s}{\pi} G^2} \, .\label{eq::LambdaE-sumrule-complete}
	\end{align}
	\end{widetext}
	\noindent

\section{Numerical Analysis} 
	\label{chp: NumericalAnalysis}
	In this section we first compute the HQET parameters by using the sum rules in Eq. \eqref{eq:SumRuleF}, (\ref{eq::LambdaHPlusE-sumrule-complete}), (\ref{eq::LambdaH-sumrule-complete}) and \eqref{eq::LambdaE-sumrule-complete} following the procedure described in Sec. \ref{chp: Contributions}. In particular, we consider the ratios \eqref{eq::LambdaHPlusE-sumrule-complete} to \eqref{eq::LambdaE-sumrule-complete} divided by \eqref{eq:SumRuleF} in order to cancel the dependence on the low-energy parameter $\bar{\Lambda}$ and the decay constant $F(\mu)$. The numerical inputs for the necessary parameters are given in Table \ref{tab::input}. But when we investigate the optimal window for the Borel parameter $M$, we observe that the sum rules are dominated by higher resonances and the continuum contribution. This questions the reliability of our estimates
	 \begin{table}[H]
	\centering
         \begin{tabular}{||c c c|} 
         \hline
         Parameters & Value & Ref. \\ [0.5ex] 
         \hline
         $\alpha_{s}$(1 GeV) & 0.471 & \cite{Herren:2017osy} \\
         \hline
         $\braket{\bar{q}q}$ & $(-0.242 \pm 0.015)^3$ GeV$^{3}$ & \cite{Jamin:2002ev} \\
        \hline
        $\braket{\frac{\alpha_{s}}{\pi} G^2}$ & $(0.012 \pm 0.004)$ \, \text{GeV}$^4$ & \cite{Gubler:2018ctz} \\
        \hline
        $\braket{\bar{q} g G \cdot \sigma q}/ \braket{\bar{q}q}$ & $(0.8 \pm 0.2)$ GeV$^2$ & \cite{Belyaev:1982sa} \\
        \hline
        $\braket{g_s^3 f^{a b c} G^{a} G^{b} G^{c}}$ & $(0.045 \pm 0.045)$ GeV$^6$ & \cite{Shifman:1978bx} \\
        \hline
        $\bar{\Lambda}$ & $(0.55 \pm 0.06)$ GeV & \cite{Gambino:2017vkx} \\
        \hline
        \end{tabular}
        \caption{List of the numerical inputs, which will be used in our analysis. The vacuum condensates are normalized at the point $\mu = 1$ GeV. For the strong coupling constant we use the two-loop expression with $\Lambda^{(4)}_{\text{QCD}} =0.31$ GeV.}
        \label{tab::input}
        \end{table}
   for $\lambda_{E,H}^2$(1 GeV) and their ratio:
	\begin{align}
	    \mathcal{R}(\mu) &=\frac{\lambda_E^2(\mu)}{\lambda_H^2(\mu)} 
	    \label{eq::R-ratio}
	\end{align} 
	at $\mu =$ 1 GeV. Hence, we study different combinations of Eq. \eqref{eq::LambdaHPlusE-sumrule-complete}, \eqref{eq::LambdaH-sumrule-complete}, \eqref{eq::LambdaE-sumrule-complete} and  \eqref{eq:SumRuleF}.
	
\begin{figure*}[hbt!]
	\centering
    \subfloat[]{\includegraphics[width=0.48 \textwidth]{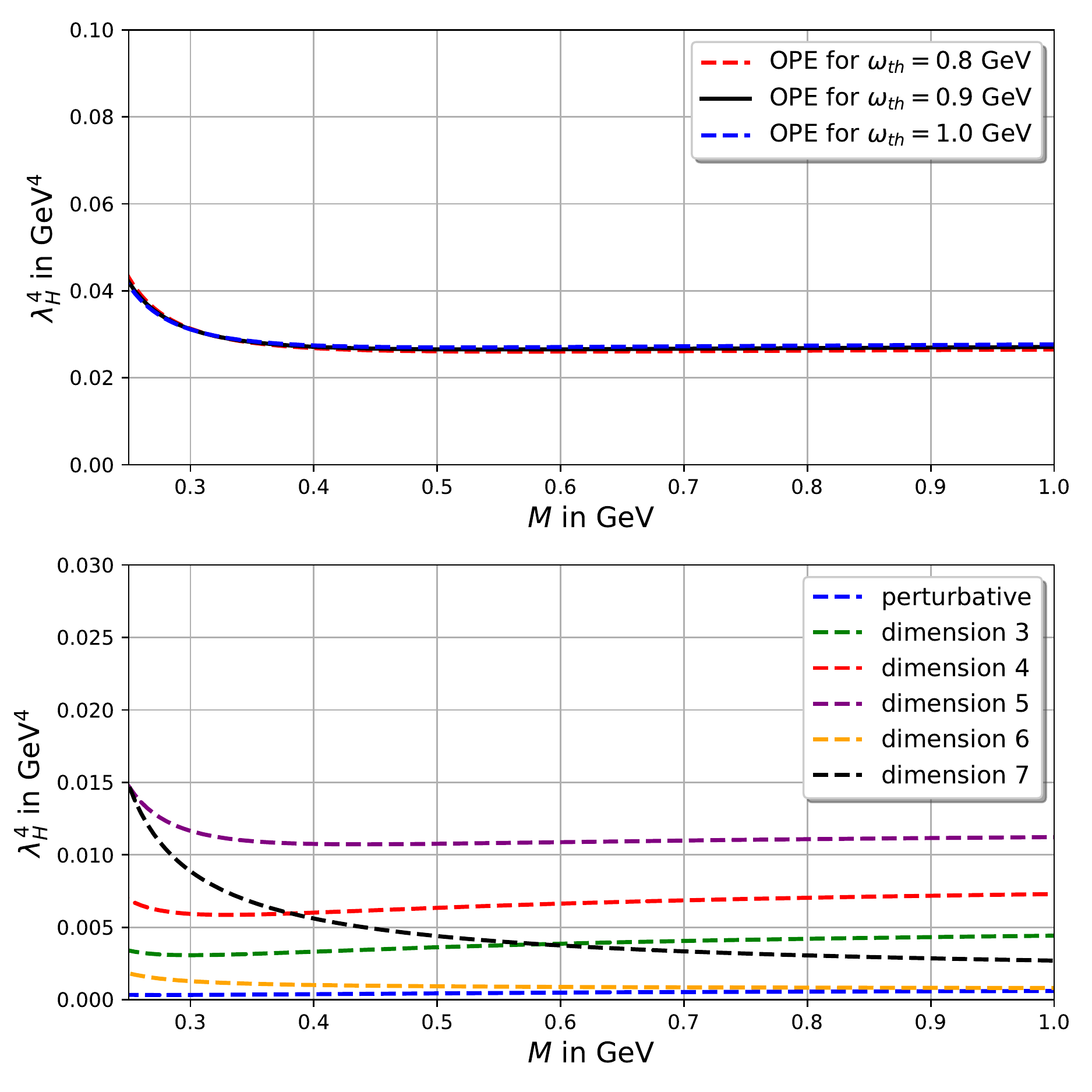}} \hspace{0.3cm}
    \subfloat[]{\includegraphics[width=0.48 \textwidth]{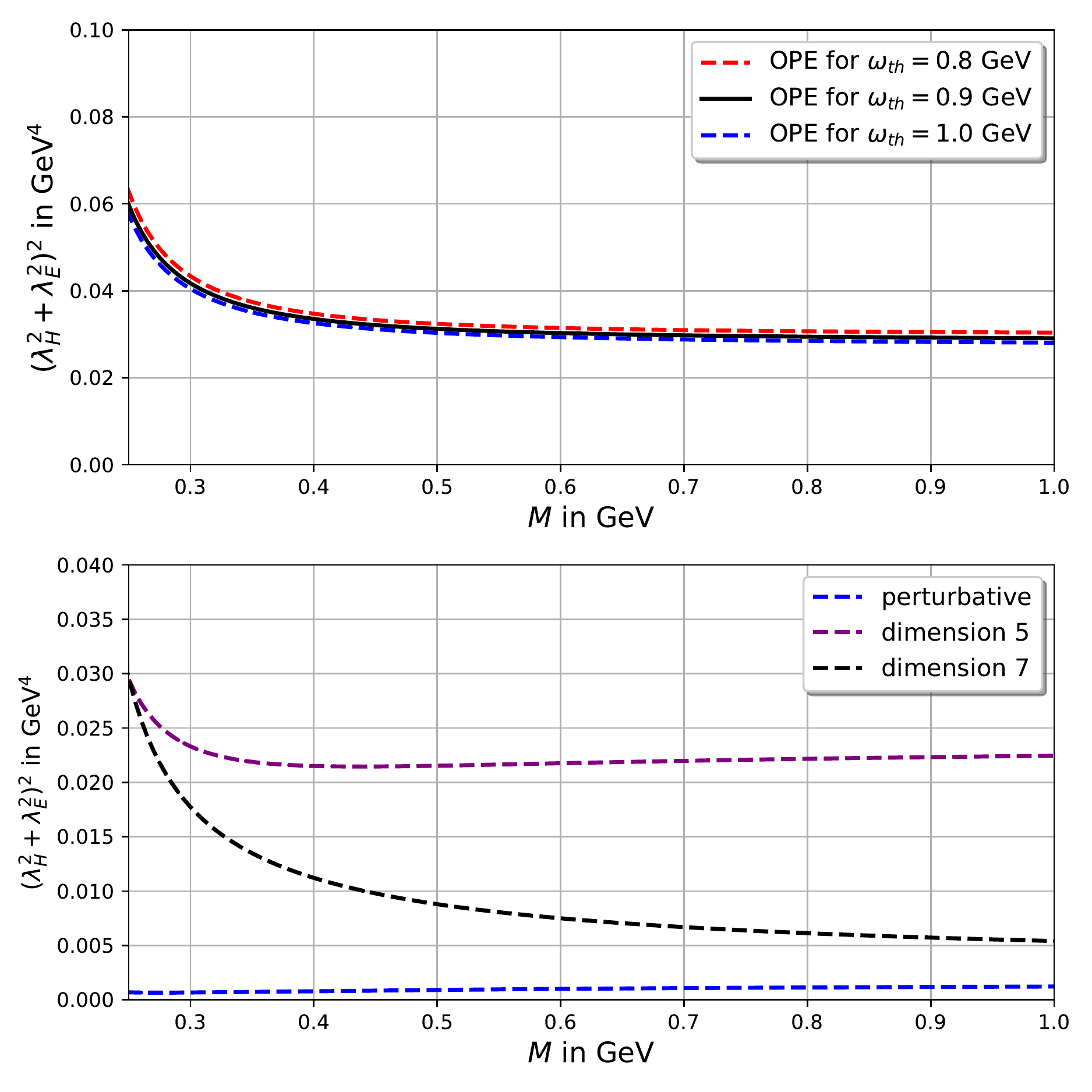}} \\
    \subfloat[]{\includegraphics[width=0.48 \textwidth]{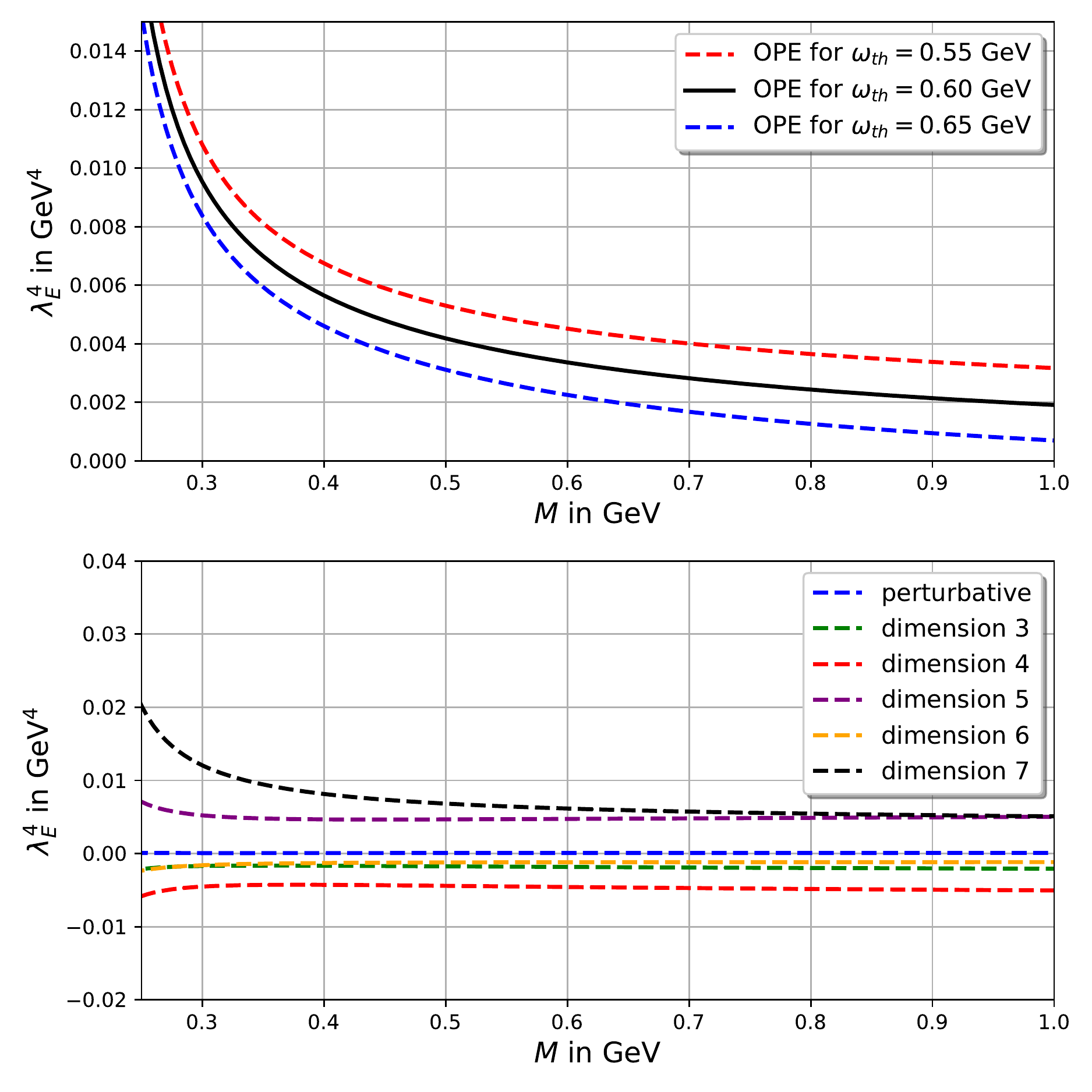}}
    \caption{Fig. (a), (b) and (c) show the full OPE of Eq. (\ref{eq::LambdaHPlusE-sumrule-complete}), (\ref{eq::LambdaH-sumrule-complete}) and (\ref{eq::LambdaE-sumrule-complete}) within the threshold interval $0.8 \, \text{GeV} \leq \omega_{th} \leq 1.0 \, \text{GeV}$, respectively. The lower figures illustrate the individual contributions to the OPE for $\omega_{th} =0.9$ GeV. The plots only show the central values.}
	\label{fig:Lambda_uptodim}
    \end{figure*}

\begin{figure*}[!htbp]
    \begin{minipage}{\linewidth}
    \centering
    \subfloat[]{\includegraphics[width=0.48 \textwidth]{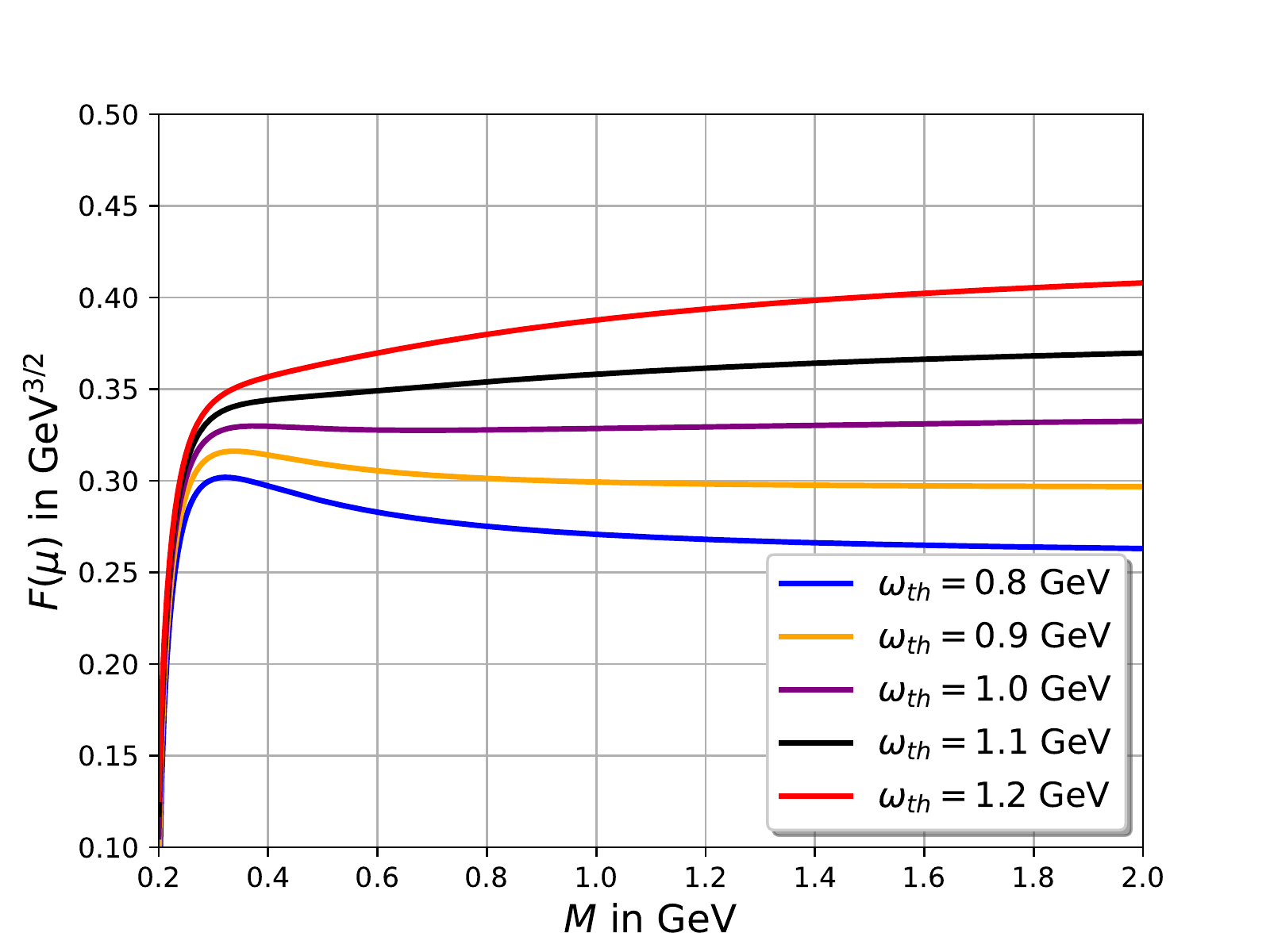}}
    \subfloat[]{\includegraphics[width=0.48 \textwidth]{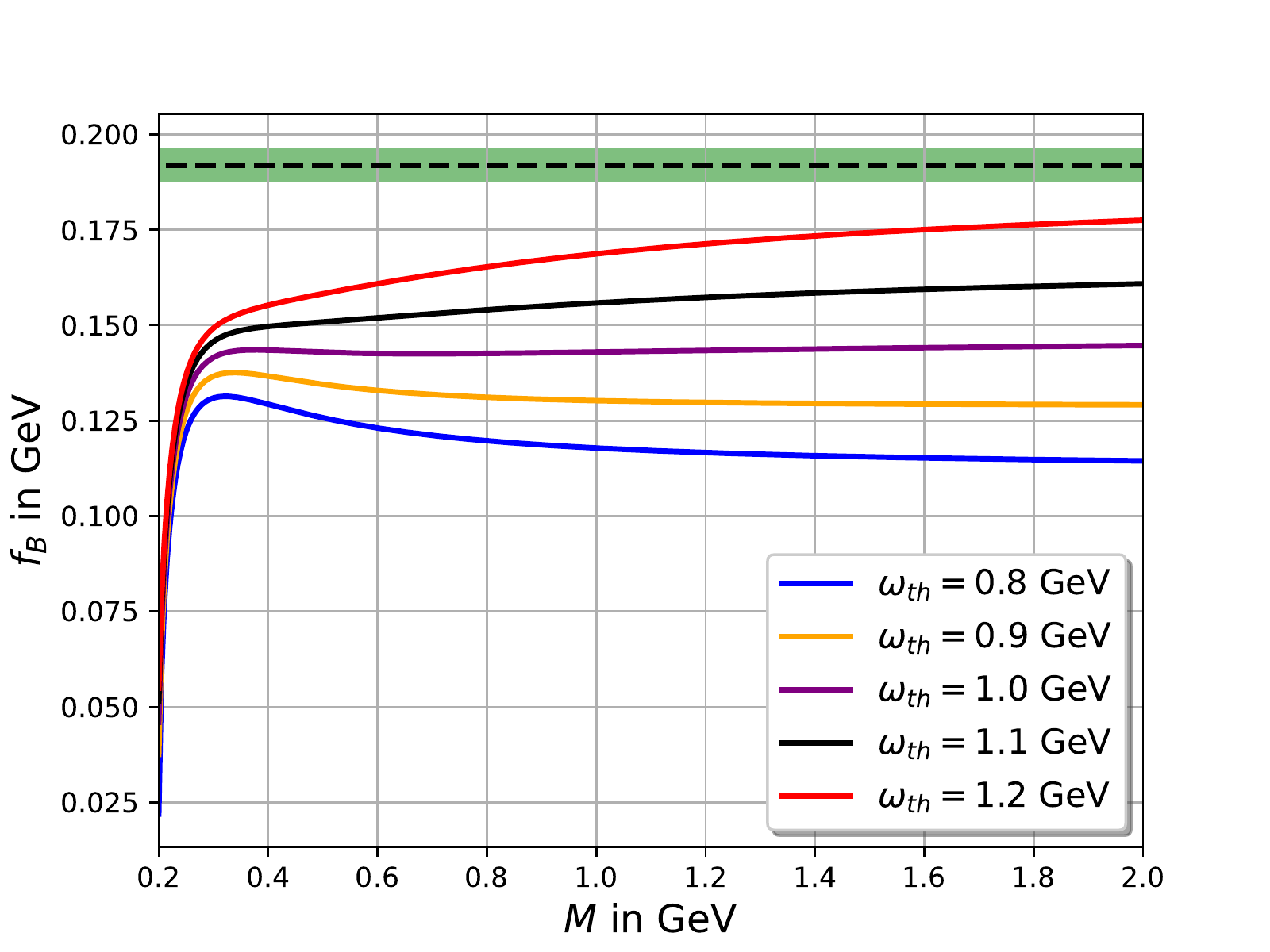}}
    \caption{Fig. (a) shows the comparison of the central values of the decay constant $F(\mu)$ for different values of $\omega_{th}$. The value of the binding energy can be found in Table \ref{tab::input}. Fig. (b) shows the comparison of the central values of the physical decay constant $f_B$ with different values of $\omega_{th}$. The dashed line indicates the lattice result and the shaded green area illustrates its corresponding uncertainty.}
    \label{fig:FalphaS}
    \end{minipage}
    \end{figure*}

    We plot higher dimensional contributions for $\lambda_{H}^4$ in the lower part of Fig. \ref{fig:Lambda_uptodim} (a) and we observe that each power correction enhances the total value of $\lambda_{H}^4$. The dimension five contribution leads to the largest contribution in Fig. \ref{fig:Lambda_uptodim} (a). The fact that correlation functions with a large mass dimension experience large contributions from local condensates with a high mass dimension for small values of the Borel parameter $M$ is a well known fact. Moreover, the contributions from dimensions greater than five become smaller indicating convergence of the OPE. The upper plot in Fig. \ref{fig:Lambda_uptodim} (a) shows the sum of all contributions up to mass dimension seven for different threshold parameters $\omega_{th}$. This variation of the parameter $\omega_{th}$ indicates the stability of the sum rule, since the Borel parameter $M$ and $\omega_{th}$ are correlated. Furthermore, it can be explicitly seen that in the highly nonperturbative regime with small $M$ the condensate contributions become dominant and therefore the sum rule becomes unreliable. To find the optimal window for the threshold $\omega_{th}$, we vary the function $F(\mu)$ in Eq. (\ref{eq:SumRuleF}) for different values of $\omega_{th}$, see Fig. \ref{fig:FalphaS} (a). As we can see, the decay constant $F(\mu)$ gives reliable values in the interval $0.8 \, \text{GeV} \leq \omega_{th} \leq 1.0 \, \text{GeV}$. In order to confirm that our threshold choice gives reasonable results, we compute the physical decay constant $f_{B}$ by using Eq. (\ref{eq:RelationFb}), see Fig. \ref{fig:FalphaS} (b). 
    We observe in Fig. \ref{fig:FalphaS} (b) that for $M \geq 0.8 \, \text{GeV}$ the dependence on the threshold parameter $\omega_{th}$ between $0.8$ GeV and $1.0$ GeV becomes stable and reliable. Although the error of the decay constant $f_B$ given in Eq. \eqref{eq::physicaldecayconstant} is small, we assume a conservative uncertainty of $50\%$, because we neglect the $\mathcal{O}(\alpha_s)$ contributions for the HQET decay constant $F(\mu)$, which are known to be large and moreover our sum rules only account for the contributions up to mass dimension seven. The corresponding analysis in \cite{Nishikawa:2011qk} shows the impact of these corrections, which reduce the uncertainty of the analysis to $15\% -20\%$. Another method to determine the interval for the threshold parameter $\omega_{th}$ is by taking the derivative with respect to the Borel parameter $\partial/\partial (-1/M)$ in Eq. \eqref{eq::LambdaH-sumrule-complete}. Dividing this expression by the original sum rule in Eq. \eqref{eq::LambdaH-sumrule-complete} yields an estimate for the parameter $\bar{\Lambda}$ which needs to be compatible with the value stated in Table \ref{tab::input}. Both methods give the same interval for $\omega_{th}$, namely $0.8 \, \text{GeV} \leq \omega_{th} \leq 1.0 \, \text{GeV}$.
    
	Similarly, we plot higher dimensional contributions for the sum rule in Eq. \eqref{eq::LambdaHPlusE-sumrule-complete} in Fig. \ref{fig:Lambda_uptodim} (b). The lower plot illustrates each order of the power expansion individually. Here, we see that the dimension three, four and six condensates do not contribute to the sum rule. The terms corresponding to the dimension five condensate provide again the largest contribution and beyond this dimension the power expansion is expected to converge, which is indicated by the small contribution of mass dimension seven. Again, the upper plot in Fig. \ref{fig:Lambda_uptodim} (b) shows the value of $(\lambda_H^2 + \lambda_{E}^2)^2$ as a function of $M$ for different threshold parameter $\omega_{th}$. The determination of the threshold window for $\omega_{th}$ follows the same argumentation as for the sum rule in Eq. \eqref{eq::LambdaH-sumrule-complete}. In particular, both methods lead again to the same conclusion and we obtain the interval $0.8 \, \text{GeV} \leq \omega_{th} \leq 1.0 \, \text{GeV}$.  
	

	The sum rule for the parameter $\lambda_E^4$ in Eq. \eqref{eq::LambdaE-sumrule-complete} requires further investigation. Fig. \ref{fig:Lambda_uptodim} (c) presents in the upper plot the sum of all contributions up to mass dimension seven, while in the lower plot each contribution is considered individually. In comparison to the sum rules in Eq. \eqref{eq::LambdaHPlusE-sumrule-complete} and Eq. \eqref{eq::LambdaH-sumrule-complete}, the mass dimension three and four condensates contribute with the opposite sign to this sum rule. Since these contributions are large, this sum rule becomes unreliable and unstable compared to the previously studied sum rules. Additionally, the dominant dimension five contributions from Eq. \eqref{eq:WilsonCoeffAbelian1}, \eqref{eq:WilsonCoeffAbelian2} and \eqref{eq:WilsonCoeffNonAbelian} do not appear in this sum rule, thus the extraction of an estimate for $\lambda_E^2$ from this sum rule gives an unreliable value. Moreover, we observe that the dimension seven contribution also gives a sizeable contribution, which questions the convergence of the OPE itself. 
	
	The fact that this sum rule becomes unstable can be seen from the threshold interval for $\omega_{th}$. Only the argumentation via the decay constants $F(\mu)$ and $f_B$ give an appropriate interval, namely $0.55 \; \text{GeV} \leq \omega_{th} \leq 0.65 \; \text{GeV}$. Furthermore, the variation of the threshold seems to give larger deviations than for the sum rules in Eq. \eqref{eq::LambdaHPlusE-sumrule-complete} and \eqref{eq::LambdaH-sumrule-complete} indicating a less stable sum rule with larger uncertainties.
    

    To obtain the lower bound for the Borel parameter $M$, we choose a value 
    where the dimension seven condensate contribution is smaller than $40 \%$ of the total OPE. Notice that too small values of $M$ spoil the convergence of the OPE since the condensate contributions become dominant. For the sum rules in Eq. (\ref{eq::LambdaHPlusE-sumrule-complete}) and (\ref{eq::LambdaH-sumrule-complete}), this condition is fulfilled for $0.5 \, \text{GeV} \leq  M$. 
    Based on Fig. \ref{fig:Lambda_uptodim} (a) and \ref{fig:Lambda_uptodim} (b), we also see that for $0.5 \, \text{GeV} \leq  M$ the sum rule starts to become more reliable. 
    As already mentioned, the sum rule for $\lambda_E^4$ in Eq. (\ref{eq::LambdaE-sumrule-complete}) is more unstable compared to $\lambda_H^4$ and $(\lambda_H^2 + \lambda_E^2)^2$.
    Hence, this method to obtain the lower bound of $M$ does not work for $\lambda_E^4$. Instead, we choose the values based on Fig. \ref{fig:Lambda_uptodim} (c). We see that for $0.5 \, \text{GeV}\leq M$ the OPE becomes more reliable and therefore a good choice for the lower bound. This estimate of the lower bound is taken into account in the uncertainty analysis.

    For the determination of the upper bound of the Borel parameter we introduce:
    \begin{align}
        R_{\text{cont.}} = 1 - \frac{\int_0^{\omega_{th}} \mathrm{d} \omega \frac{1}{\pi} \mathrm{Im} \Pi_X^{\text{OPE}}(\omega) e^{-\omega/M}}{\int_0^{\infty} \mathrm{d} \omega \frac{1}{\pi} \mathrm{Im} \Pi_X^{\text{OPE}}(\omega) e^{-\omega/M}} 
        \label{eq::R-continuum}
    \end{align}
    for $X \in \{H,E,HE\}$ .
    The value of $R_{\text{cont.}}$ guarantees that the ground state still gives a sizeable contribution compared to the higher resonances and continuum contribution.
    For reliable results of the sum rule we expect $R_{\text{cont.}} \leq 50 \%$ for $M \leq M_{\text{max}}$. 
    Thus, Eq. (\ref{eq::R-continuum}) fixes the upper bound for the Borel parameter. But in the case of Eq. \eqref{eq::LambdaHPlusE-sumrule-complete}, \eqref{eq::LambdaH-sumrule-complete} and \eqref{eq::LambdaE-sumrule-complete}, the continuum contribution is dominant, which is to be expected from the large mass dimension of the considered correlation function in Eq. \eqref{eq:CorrelationFunc}. Therefore, an upper bound for $M$ is not feasible according to this method. 
    
    To resolve this problem, we consider two combinations of the sum rules in Sec. \ref{chp: Contributions}, which have the feature that $R_{\text{cont.}}$ becomes about $50 \%$ for a reasonable value of $M$. The combinations are the following:
    \begin{align}
        &\frac{(\lambda_H^2 + \lambda_E^2)^2 }{\lambda_H^4} = (1 + \mathcal{R})^2 \hspace{0.5cm} \text{and} \nonumber \\& \frac{F(\mu)^2 e^{-\bar{\Lambda}/M} + F(\mu)^2 e^{-\bar{\Lambda}/M} \lambda_H^4}{F(\mu)^2 e^{-\bar{\Lambda}/M} - F(\mu)^2 e^{-\bar{\Lambda}/M} \lambda_E^4} \label{eq:Comb1}
    \end{align}
    with $\mathcal{R}$ defined in Eq. (\ref{eq::R-ratio}). The combination $(1+\mathcal{R})^2$ is an appropriate choice, because the dominant mass dimension five contributions due to Eq. \eqref{eq::LambdaH-sumrule-complete} lower the value of $R_{\text{cont.}}$ significantly. 
    On the other hand, the second combination in Eq. (\ref{eq:Comb1}) is dominated by the large $\mathcal{O}(\alpha_s^0)$ contributions from $F(\mu)$ such that $\lambda_{E,H}^4$ become only small corrections. 
    For both combinations in Eq. (\ref{eq:Comb1}) the parameter is $R_{\text{cont.}} \leq 50 \%$ for $M_{\text{max}} = 0.8$ GeV. 
    
\begin{figure*}[t]
	\centering
   \subfloat[]{\includegraphics[width=0.48 \textwidth]{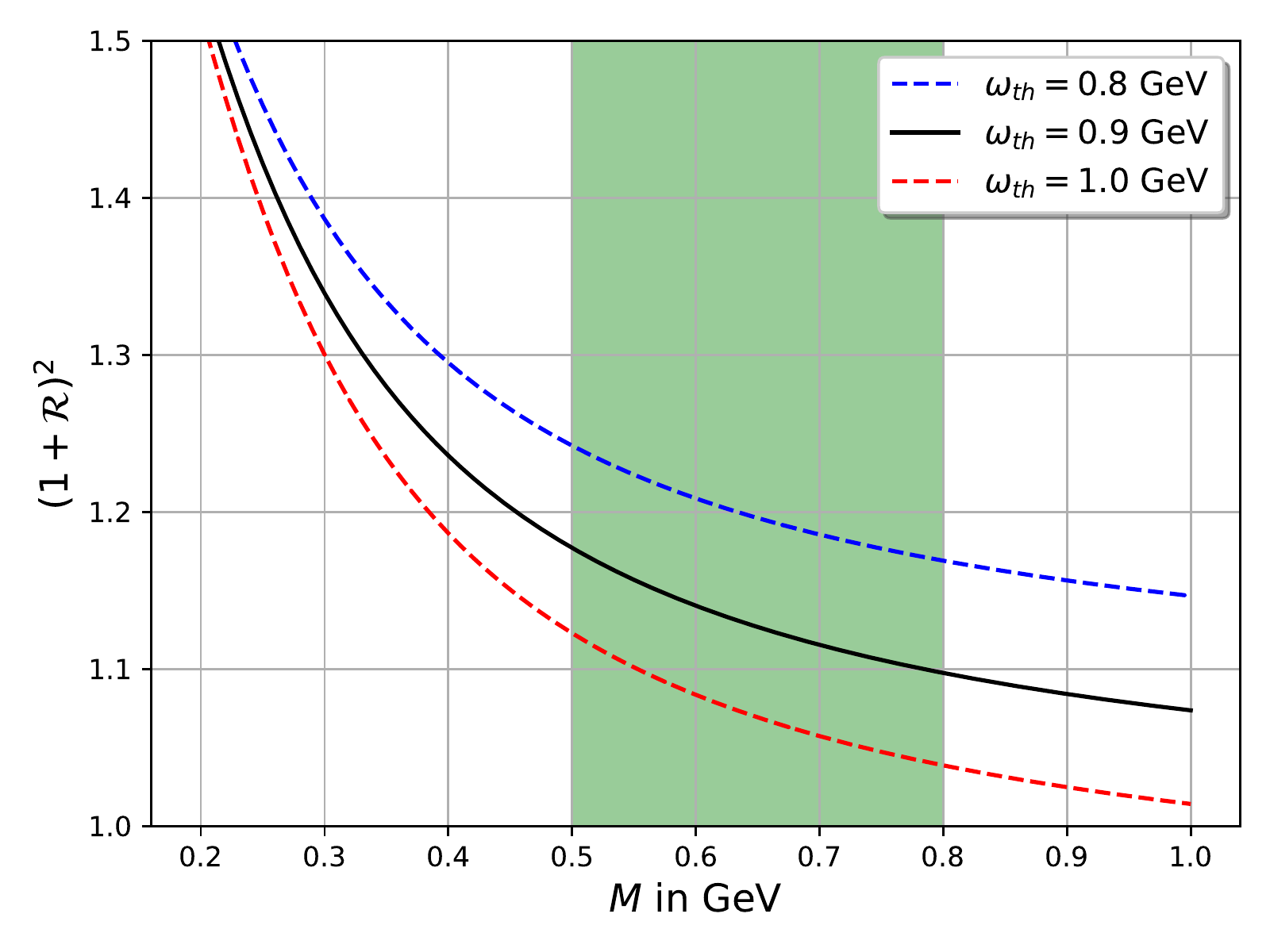}}\hspace{0.3cm}
   \subfloat[]{ \includegraphics[width=0.48 \textwidth]{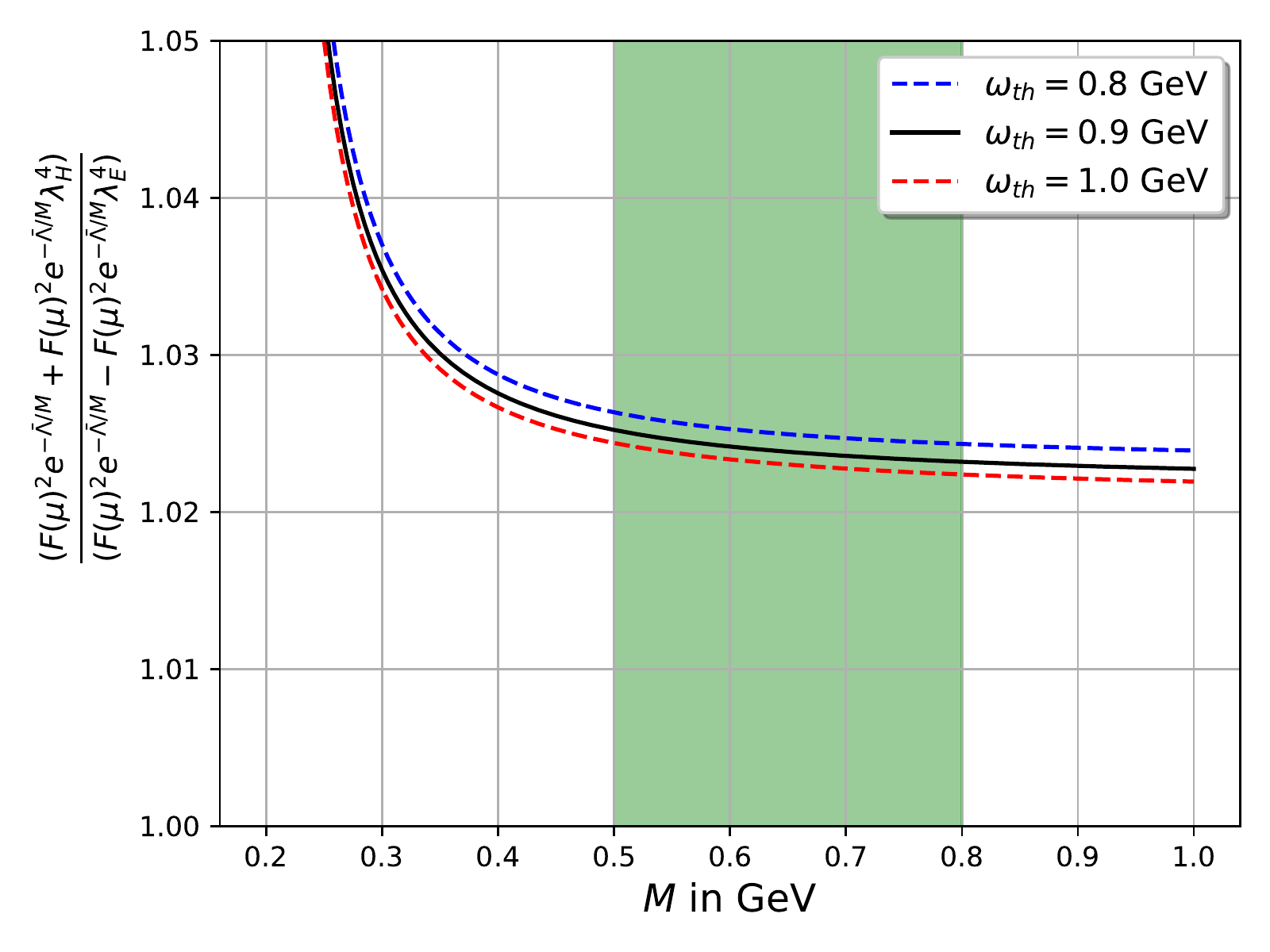}}
	\caption{Fig. (a) shows the Borel sum rule for $(1+\mathcal{R})^2$ for the window $0.8 \, \text{GeV} \leq \omega_{th} \leq 1.0 \, \text{GeV}$. The shaded green area illustrates the Borel window. Similarly, Fig. (b) shows the Borel sum rule for $(F(\mu)^2 e^{- \bar{\Lambda}/M} + F(\mu)^2 e^{- \bar{\Lambda}/M} \lambda_H^4)/(F(\mu)^2 e^{- \bar{\Lambda}/M} - F(\mu)^2 e^{- \bar{\Lambda}/M} \lambda_E^4)$ for the window $0.8 \, \text{GeV} \leq \omega_{th} \leq 1.0 \, \text{GeV}$.}
	\label{fig::Comb1}
    \end{figure*}

    In Table \ref{tab::ThresholdAndBorel} we summarize the lower and upper bounds for the parameters $M$ and $\omega_{th}$. 
    \begin{table}[H]
	\centering
	\scalebox{0.95}{
         \begin{tabular}{||c c c|} 
         \hline
          Sum rule & Borel window & threshold window \\ [0.5ex] 
         \hline
         Eq. (\ref{eq:Comb1}) & $0.5 \; \mathrm{GeV} \leq M \leq 0.8 \; \mathrm{GeV}$ & $0.8 \; \mathrm{GeV} \leq \omega_{th} \leq 1.0 \; \mathrm{GeV}$ \\
        \hline
        \end{tabular}}
        \caption{Summary of the threshold and Borel window for the combination in Eq. (\ref{eq:Comb1}).}
        \label{tab::ThresholdAndBorel}
    \end{table}
    In Fig. \ref{fig::Comb1} (a) and \ref{fig::Comb1} (b) we plot both combinations as a function of $M$ for different values of $\omega_{th}$ within its threshold window. 
    
    Finally, we are at the point to extract $\mathcal{R}$ and $\lambda_{E,H}^2$ based on Eq. (\ref{eq:Comb1}). The uncertainties of $\lambda_{E,H}^2$ and for the ratio $\mathcal{R}$ are partially determined by varying each input parameter individually according to their uncertainty, see Table \ref{tab::input}. For the strong coupling constant we use the two-loop expression with $\Lambda_{\text{QCD}}^{(4)} = 0.31$ GeV to obtain $\alpha_s(1 \, \text{GeV}) = 0.471$. We vary $\Lambda_{\text{QCD}}^{(4)}$ in the interval $0.29 \, \text{GeV} \leq \Lambda_{\text{QCD}}^{(4)} \leq 0.33 \, \text{GeV}$, which corresponds to the running coupling $\alpha_s(1 \, \text{GeV}) = 0.44 - 0.50$. In the last step, we square each uncertainty in quadrature: 
    \begin{align}
        \mathcal{R}(1 \, \text{GeV}) &=  0.1 + \left(\begin{array}{c}+ 0.03\\-0.03\\\end{array}\right)_{\omega_{th}} +\left(\begin{array}{c}+ 0.01\\-0.02\\\end{array}\right)_{M}  \nonumber \\
        & \hspace{-0.35cm}+ \left(\begin{array}{c} + 0.01\\-0.01\\\end{array}\right)_{\alpha_s} +  \left(\begin{array}{c}+ 0.01\\-0.01\\\end{array}\right)_{\braket{\bar{q} q}}  + \left(\begin{array}{c}+ 0.02\\-0.03\\\end{array}\right)_{\braket{\frac{\alpha_{s}}{\pi} G^2}}  \nonumber \\
        & \hspace{-0.35cm} + \left(\begin{array}{c}+ 0.05\\-0.04\\\end{array}\right)_{\braket{\bar{q} g G \cdot \sigma q}} + \left(\begin{array}{c}+ 0.02\\-0.02\\\end{array}\right)_{\braket{g_s^3 f^{a b c} G^{a} G^{b} G^{c}}} \nonumber \\
        &= 0.1 \pm 0.07  \label{eq::R_final}
    \end{align}
    \begin{align}
        \lambda_H^2(1 \, \text{GeV})  &= \Big [ 0.150 + \left(\begin{array}{c}+ 0.002\\-0.003\\\end{array}\right)_{\omega_{th}} +\left(\begin{array}{c}+ 0.002\\-0.004\\\end{array}\right)_{M} \nonumber \\
        & + \left(\begin{array}{c}+ 0.001\\-0.001\\\end{array}\right)_{\braket{\frac{\alpha_{s}}{\pi} G^2}} + \left(\begin{array}{c}+ 0.001\\-0.001\\\end{array}\right)_{\braket{\bar{q} g G \cdot \sigma q}} \nonumber \\ 
        &  + \left(\begin{array}{c}+ 0.001\\-0.001\\\end{array}\right)_{\braket{g_s^3 f^{a b c} G^{a} G^{b} G^{c}}} \Big] \, \text{GeV}^2 \nonumber \\
        &= (0.150 \pm 0.006) \,\text{GeV}^2 
        \label{eq::lambdaH_final}
    \end{align}
    
    For $\lambda_H^2$, the variation of the strong coupling constant $\alpha_s$, the dimension three and dimension six condensates do not change the central value significantly. Therefore, these uncertainties can be neglected.
    \begin{align}
        \lambda_E^2(1 \, \text{GeV}) &= \Big [ 0.010 + \left(\begin{array}{c}+ 0.004\\-0.005\\\end{array}\right)_{\omega_{th}} +\left(\begin{array}{c}+ 0.002\\-0.003\\\end{array}\right)_{M}  \nonumber \\
        & + \left(\begin{array}{c}+ 0.001\\-0.001\\\end{array}\right)_{\alpha_s} + \left(\begin{array}{c}+ 0.003\\-0.003\\\end{array}\right)_{\braket{\bar{q} q}}  \nonumber \\ 
        &  + \left(\begin{array}{c}+ 0.003\\-0.004\\\end{array}\right)_{\braket{\frac{\alpha_{s}}{\pi} G^2}} +\left(\begin{array}{c}+ 0.007\\-0.006\\\end{array}\right)_{\braket{\bar{q} g G \cdot \sigma q}} \nonumber \\
        & + \left(\begin{array}{c}+ 0.002\\-0.002\\\end{array}\right)_{\braket{g_s^3 f^{a b c} G^{a} G^{b} G^{c}}} \Big ] \, \text{GeV}^2 \nonumber \\
        &= (0.010 \pm 0.009) \,\text{GeV}^2 \, .
        \label{eq::lambdaE_final}
    \end{align}

   	Notice that the threshold parameter $\omega_{th}$ and the Borel parameter $M$ are correlated, which can be deduced from the determination of the Borel window and the threshold interval. But since the variation of $\omega_{th}$ with respect to $M$ is negligible, it is possible to choose one point in the parameter space of both parameters where the conditions from above are satisfied and obtain an estimate for the uncertainty by varying $\omega_{th}$.
   	
   	Besides these contributions, there are other uncertainties due to several approximations and systematic errors. Since we truncated the perturbative series at $\mathcal{O}(\alpha_s)$ and the power corrections at dimension seven, we introduce another error which is more complicated to determine. Moreover, there is also an intrinsic uncertainty caused by the sum rule approach, for instance generated by the use of the quark-hadron duality. The total uncertainties stated in Eq. \eqref{eq::R_final}, \eqref{eq::lambdaH_final} and \eqref{eq::lambdaE_final} only list those quantities, which give deviations from the central values. 
   	
   	Before we state our final results, we will first derive upper bounds on the parameters $\lambda_{E,H}^2$. Due to the diagonal structure of the correlation function, we know that the spectral density is positive definite. By performing the limit $\omega_{th} \rightarrow \infty$ in Eq. \eqref{eq::LambdaH-sumrule-complete} and \eqref{eq::LambdaE-sumrule-complete}, we include all possible higher resonances and continuum contributions into our analysis. Thus, we obtain a consistent upper bound onto these parameters as it was already done in the case of $f_D/f_{D_s}$ decay constants in \cite{Khodjamirian:2008xt}. The values for the upper bounds within the Borel window in Fig. \ref{fig::Comb1} (a) and \ref{fig::Comb1} (b) are:
	\begin{align}
	    \lambda_H^2 &< 0.48^{+ 0.17}_{-0.24} \, \; \text{GeV}^2 \, , \label{eq:UpperBoundH}\\
	    \lambda_E^2 &< 0.41^{+ 0.19}_{-0.24} \, \; \text{GeV}^2 \, . \label{eq:UpperBoundE}
	\end{align}
   	Now we extract our predictions for these parameters based on our sum rule analysis. We expect that these estimates should lie within the bounds of \eqref{eq:UpperBoundH} and \eqref{eq:UpperBoundE}.
   	A conservative estimate of the uncertainties leads to the following final results:
	\begin{align}
	    \lambda_{E}^2(\text{1 GeV}) &= (0.01 \pm 0.01) \, \, \text{GeV}^2 \, , \label{eq:EstimateE} \\
	    \lambda_{H}^2(\text{1 GeV}) &= (0.15 \pm 0.05) \, \, \text{GeV}^2 \, , \label{eq:EstimateH} \\
	    \mathcal{R} &= 0.1 \pm 0.1 \, . \label{eq:EstimateR}
	\end{align}
	If we consider instead directly Eq. \eqref{eq::LambdaHPlusE-sumrule-complete}, \eqref{eq::LambdaH-sumrule-complete} and take the Borel window and the threshold parameter $\omega_{th}$ as shown in Table \ref{tab::ThresholdAndBorel},  we obtain the values:
	\begin{align}
	    \lambda_{E}^2(\text{1 GeV}) &= (0.05 \pm 0.03) \, \, \text{GeV}^2 \, , \label{eq:EstimateEContProblem}\\
	    \lambda_{H}^2(\text{1 GeV}) &= (0.16 \pm 0.05) \, \, \text{GeV}^2 \, , \label{eq:EstimateHContProblem}\\
	    \mathcal{R} &= 0.3 \pm 0.2 \, . \label{eq:EstimateRContProblem}
	\end{align}
	Note that we can also use \eqref{eq::LambdaE-sumrule-complete} to obtain the value for $\lambda_E^2$, however the threshold window must be chosen as $0.55 \, \text{GeV} \leq \omega_{th} \leq 0.65 \, \text{GeV}$ as shown in Fig. \ref{fig:Lambda_uptodim} (c).

    Although the sum rules in Eq. \eqref{eq::LambdaHPlusE-sumrule-complete} to \eqref{eq::LambdaE-sumrule-complete} are dominated by continuum contributions and higher resonances for the Borel window given in Table \ref{tab::ThresholdAndBorel}, we see that the set of parameters and their ratio $\mathcal{R}$ in Eq. \eqref{eq:EstimateEContProblem} to \eqref{eq:EstimateRContProblem} reproduce the values for $\lambda_{E,H}^2$ and $\mathcal{R}$ in Eq. \eqref{eq:EstimateE} to \eqref{eq:EstimateR} within the errors. In particular the estimate for $\lambda_H^2$ does not change much, which indicates that the continuum contributions are well approximated by the sum rules in Eq. \eqref{eq::LambdaH-sumrule-complete}. All values lie within the bounds given in Eq. \eqref{eq:UpperBoundH} and \eqref{eq:UpperBoundE}.
    Our result for $\lambda_{E}^2$ in Eq. \eqref{eq:EstimateE} is close to the result in \cite{Nishikawa:2011qk} and agrees within the error, see Table \ref{tab::finalresult}. Additionally, our result for $\lambda_{H}^2$ tends towards the result in \cite{Grozin:1996pq}. 
    

    \section{Conclusion}
    \label{chp:Conclusion}
    In this work we suggested alternative diagonal QCD sum rules in order to estimate the HQET parameters $\lambda_{E,H}^2$ and their ratio $\mathcal{R} = \lambda_E^2/\lambda_H^2$. We included all leading contributions to the diagonal correlation function of three-particle quark-antiquark-gluon currents up to mass dimension seven. The advantage of these sum rules are that they are positive definite and we expect that the quark-hadron duality is more accurate compared to the previously studied correlation functions in \cite{Grozin:1996pq,Nishikawa:2011qk}. But 
    we observe dominant contributions from the continuum and higher resonances due to the large mass dimension of the correlation function within these sum rules. This is why we consider combinations of these sum rules studied in Section \ref{chp: NumericalAnalysis}, which satisfy the condition that the ground state contribution still gives a sizeable effect. Moreover, the OPE is expected to converge for the two sum rules in Eq. (\ref{eq::LambdaHPlusE-sumrule-complete}) and (\ref{eq::LambdaH-sumrule-complete}) shown in Fig. \ref{fig:Lambda_uptodim} (a) and \ref{fig:Lambda_uptodim} (b), because the investigated contributions beyond mass dimension five become smaller. 
    However, the OPE in Eq. (\ref{eq::LambdaE-sumrule-complete}) needs additional higher order corrections, since the contribution of dimension five and seven are both large, which makes the sum rule unstable, see Fig. \ref{fig:Lambda_uptodim} (c). 
    
    For a consistent treatment of the leading order contributions we also included only the $\mathcal{O}(\alpha_s^0)$ contributions for the HQET decay constant $F(\mu)$, although it is known that the $\mathcal{O}(\alpha_s)$ contributions are sizeable \cite{Broadhurst:1991fc}.
    Our results compared to the values obtained in \cite{Grozin:1996pq,Nishikawa:2011qk} are listed in Table \ref{tab::finalresult}.
    \begin{table}[H]
	\centering
	\scalebox{0.80}{
         \begin{tabular}{||c c c c|} 
         \hline
         Parameters & Ref. \cite{Grozin:1996pq}  & Ref. \cite{Nishikawa:2011qk}  & \textbf{this work} \\ [0.5ex] 
         \hline
        $\mathcal{R}$(1 GeV) &  (0.6 $\pm$ 0.4)  &  (0.5 $\pm$ 0.4) & (0.1 $\pm$ 0.1) \\
        $\lambda_H^2$(1 GeV) & (0.18 $\pm$ 0.07) GeV$^2$ & (0.06 $\pm$ 0.03) GeV$^2$  & (0.15 $\pm$ 0.05) GeV$^2$ \\
         $\lambda_E^2$(1 GeV) & (0.11 $\pm$ 0.06) GeV$^2$ & (0.03 $\pm$ 0.02) GeV$^2$ &(0.01 $\pm$ 0.01) GeV$^2$ \\
        \hline
        \end{tabular}}
        \caption{Comparison of our results for the parameters $\lambda_{E,H}^2$ and $\mathcal{R}$ at $\mu = 1 \; \mathrm{GeV}$.}
        \label{tab::finalresult}
        \end{table}
 
    With these new sum rules we obtain independent estimates for the parameters $\lambda_{E,H}^2$ and the $\mathcal{R}$-ratio, which are important ingredients for the second moments of the $B$-meson light-cone distribution amplitudes in $B$-meson factorization theorems. 
    For future improvements of our sum rules we suggest to include $\mathcal{O}(\alpha_s^2)$ corrections to the OPE and consider even higher mass dimension in the power expansion of local vacuum condensates. In this case it would also be necessary to include the $\mathcal{O}(\alpha_s)$ contributions for $F(\mu)$. Especially the sum rule in \eqref{eq::LambdaE-sumrule-complete} will benefit greatly since we expect the convergence of the OPE, which results in better determination of $\lambda_{E,H}^2$ and consequently $\mathcal{R}$.

\acknowledgments
   We would like to thank Alexander Khodjamirian for proposing this project to us, for his constant feedback throughout the work and for reading the manuscript. We thank Thomas Mannel for useful discussion and reading the manuscript. Additionally, we are grateful to Thorsten Feldmann and Alexei Pivovarov for helpful discussions. This research was supported by the Deutsche Forschungsgemeinschaft (DFG, German Research Foundation) under grant  396021762 - TRR 257.


\appendix

\section{PARAMETRIZATION OF THE QCD CONDENSATES} \label{chp:Condensate}
	Here we present the condensates that we have used in the work. All results are based on \cite{Pascual:1984zb} if not stated otherwise.
	\noindent
	First, we Taylor expand the following matrix element:
	\begin{small}
	\begin{align}
	    \bra{0} \bar{q}(0) \Gamma_{1} P_{+} \Gamma_{2} \,q(x) \ket{0} &= \bra{0} \bar{q}(0) \Gamma_{1} P_{+} \Gamma_{2} q(0) \ket{0} \nonumber \\& + x^{\mu} \bra{0} \bar{q}(0) \Gamma_{1} P_{+} \Gamma_{2} D_{\mu} q(0) \ket{0} \nonumber \\
	    & + \frac{x^{\mu} x^{\nu}}{2}  \bra{0} \bar{q}(0) \Gamma_{1} P_{+} \Gamma_{2} D_{\mu} D_{\nu} q(0) \ket{0} \nonumber \\& + \cdots
	    \label{eq::matrix25}
	\end{align}
	\end{small}
    The first term in Eq.(\ref{eq::matrix25}) corresponds to the quark condensate. 
    \begin{align}
	    \bra{0} \bar{q}^{i}_{\alpha}(0) \Gamma_{1,\alpha \beta} P_{+, \beta\gamma} \Gamma_{2, \gamma \delta} \,q^{j}_{\delta}(0) \ket{0}  =& \; \frac{1}{4  N_c} \cdot \text{Tr}[\Gamma_{1} P_{+} \Gamma_{2}] \nonumber \\
	    & \times \braket{\bar{q} q} \delta^{ij},
    \end{align}
    where $(i,j)$ are color indices and $(\alpha, \beta,\gamma,\delta)$ are spinor indices. 
    \noindent
    The second term in Eq.(\ref{eq::matrix25}) does not contribute. Making use of the Dirac equation, we can rewrite the covariant derivative as:
    \begin{align}
        \slashed{D} q = - i m_{q} q \, .
    \end{align}
    We assume $m_{q} = 0$ for light quarks.
    
    Before we consider the third term in more detail, we parametrize the dimension five matrix element:
    \begin{align}
	    \bra{0} \bar{q}_{\alpha}^i(0)g_s G_{\mu \nu}(0) q_{\delta}^j(0) \ket{0} =& \; \bra{0} \bar{q} g_s \sigma \cdot G q \ket{0} \cdot \frac{1}{4 N_c d (d - 1)}  \nonumber \\
	    & \times \delta^{ij} \cdot (\sigma_{\mu \nu})_{\delta \alpha} \, .
    \end{align}
    The third term in Eq.(\ref{eq::matrix25}) corresponds to the quark-gluon condensate.
	\begin{align}
		\frac{x^{\mu} x^{\nu}}{2} \bra{0}\bar{q}^{i}_{\alpha}(0) D_{\mu} D_{\nu} q^{j}_{\delta}(0) \ket{0} =& \frac{x^2}{16 N_c d} \, \delta^{ij} \delta_{\alpha \delta} \bra{0} \bar{q} g_s \sigma \cdot G q \ket{0} \, .
	\end{align}
   The gluon condensate can be parametrized as:
   \begin{align}
       \bra{0} G_{\mu \nu}^{a} G_{\rho \sigma}^{b} \ket{0} &= \frac{\delta^{ab}}{d (d-1) (N_c^2-1)} \braket{G^2} (g_{\mu \rho} g_{\nu \sigma} - g_{\mu \sigma} g_{\nu \rho}) \, .
   \end{align}
   
	Next is the parametrization of the triple-gluon condensate, which was denoted as $B_{\mu \lambda \rho \nu \sigma \alpha}$ in Eq. \eqref{eq:tripleGluon}. The decomposition of the triple-gluon condensate has been investigated in ~\cite{Nikolaev:1982rq}:
	\begin{widetext}
	\begin{align}
	    \braket{g_s^3 f^{abc} G^a_{\mu \nu} G^b_{\rho \sigma} G^c_{\alpha \lambda}} =& \; \frac{\braket{g_s^3 f^{abc} G^a G^b G^c}}{d (d - 1) (d - 2)} \cdot \Big(g_{\mu \lambda} g_{\rho \nu} g_{\sigma \alpha} + g_{\mu \sigma} g_{\rho \alpha} g_{\lambda \nu} + g_{\rho \lambda} g_{\mu \alpha} g_{\nu \sigma} + g_{\alpha \nu} g_{\mu \rho} g_{\sigma \lambda} -  \nonumber \\& \; g_{\mu \sigma} g_{\rho \lambda} g_{\alpha \nu} - g_{\mu \lambda} g_{\rho \alpha} g_{\nu \sigma} - g_{\rho \nu} g_{\mu \alpha} g_{\sigma \lambda} - g_{\sigma \alpha} g_{\mu \rho} g_{\nu \lambda} \Big)  \, .
	    \label{eq:DecompDim6}
	\end{align}
	\end{widetext}
	\noindent
	The expression in Eq. \eqref{eq:DecompDim6} corresponds to the tensor $B_{\mu \lambda \rho \nu \sigma \alpha}$ introduced in Eq. \eqref{eq:tripleGluon}.


\bibliographystyle{apsrev4-2}
\bibliography{main}

\end{document}